\documentclass[twocolumn,aps,pra,showpacs,groupedaddress]{revtex4}
\usepackage{epsfig,color} 
\usepackage{amsmath,amssymb,mathtools,amsthm,mathrsfs,stmaryrd}
\usepackage[linkcolor=red]{hyperref}

\def\syst#1{\mathrm #1}
\newcommand\rA{\syst A}
\newcommand\nats{\mathbb N}
\def\bvec#1{\mathbf{#1}}
\def\transf#1{\mathscr{#1}}
\def\tT{\transf T}
\def\tZ{\transf Z}

\usepackage{wrapfig}
\begin{document}

\title{Scalar fermionic cellular automata on finite Cayley graphs}
\author{Paolo Perinotti}\email{paolo.perinotti@unipv.it} 
\affiliation{{\em QUIT Group}, Dipartimento di Fisica, Universit\`a degli studi di Pavia, and INFN sezione di Pavia, via Bassi 6, 27100 Pavia, Italy}
\homepage{http://www.qubit.it}
\author{Leopoldo Poggiali}\email{leopoldo.poggiali01@ateneopv.it}
\affiliation{{\em QUIT Group}, Dipartimento di Fisica, Universit\`a degli studi di Pavia, and INFN sezione di Pavia, via Bassi 6, 27100 Pavia, Italy}
\homepage{http://www.qubit.it}

\date{\today}

\begin{abstract}
A map on finitely many fermionic modes represents a unitary evolution if and only if it preserves canonical anti-commutation relations. We use this condition for the classification of fermionic cellular automata (FCA) on Cayley graphs of finite groups in two simple but paradigmatic case studies. The physical properties of the solutions are discussed. Finally, features of the solutions that can be extended to the case of cellular automata on infinite graphs are analyzed. 
\end{abstract}
\maketitle

\section{Introduction}
The experience of quantum information and computation theory \cite{Nielsen:2000aa} represented a cornerstone for the progress of the understanding of the structure of quantum theory, and refreshed the traditional approaches to quantum foundations, based on the vision of quantum theory as a special theory of information processing \cite{Hardy:2001aa,Fuchs:2002aa,Brassard:2005aa,DAriano:2010aa}. This approach culminated a decade later in a wealth of reconstructions of quantum theory \cite{Dakic09,Masanes10,Hardy2016}, among which a fully informational axiomatisation of the mathematical framework of the theory was obtained in Ref.~\cite{Chiribella:2011aa} (see also Ref.~\cite{Chiribella17}).

The understanding of quantum theory as a theory of information processing brought about the question as whether it is possible to derive the full quantum mechanics, including dynamical laws such as e.g.~Dirac's equation, from a purely informational background, rephrasing a problem that in a slightly different form was posed in one of the seminal papers on quantum computation by Feynman \cite{Feynman82}. A recent successful approach to the above question allowed for a reconstruction of the geometry of Minkowski spacetime and the free dynamics of relativistic quantum fields \cite{DAriano:2014aa}. This approach is based on the notion of Fermionic Cellular Automata (FCAs), namely cellular automata on local Fermionic systems~\cite{BravyiKitaev02,DarianoetAl14,DarianoManessi14}. 

The main results presently achieved in the above context are the derivation of Weyl's and Dirac's equations in 1+1, 2+1, and 3+1 dimensions \cite{BISIO2015244,Bisio:2013aa,DAriano:2014aa}, Maxwell's equations with a suitable approximation of the Bosonic statistics \cite{Bisio16}, along with their symmetry under realisations of the Poicar\'e group \cite{Bibeau-Delisle15,Bisio:2016aa,Bisio:2016ab,Bisio2017}. All the automata studied so far are linear, namely they evolve field operators into linear combinations of field operators. From a technical point of view, linear FCAs are closely related to the literature on Lattice Gas Automata and Quantum Walks \cite{Nakamura:1991aa,Bialynicki-Birula94,Meyer96,Yepez05,2018arXiv180203910M,Brun:2018wo}. Linear FCAs can describe free fields, without interactions. There are very few exceptions, where non-linear FCA have been studied \cite{Bisio:2018aa,Bisio:2018ab}, describing non trivial interactions. Non-linear unitary FCAs essentially remain an unknown subject. The study of non-linear FCAs, however, is a very relevant matter in the reconstruction program, as non-linearity of the evolution is a necessary condition for the expression of non trivial interacting theories.

The mathematical definition of a Quantum Cellular Automaton (QCA) was provided in Ref.~\cite{Schumacher:2004aa}. Here we generalize the definition in the case of FCA, and show that the analogue of the so-called {\em wrapping lemma} holds, that allows one to reduce the evolution of infinite FCAs to that of finite-dimensional ones. Using the construction of Ref.~\cite{Nielsen05}, we know that a necessary and sufficient condition for a unitarity evolution of FCAs is the preservation of the Fermionic algebra, defined by the canonical anti-commutation relations.

The relevance of finite-dimensional FCAs is then much broader than one could expect, providing sufficient information for the classification and understanding of infinite FCAs as well. The purpose of the present analysis is to figure out conditions that underpin unitarity in the special case studies, with the final objective of finding general features extendable to the general case. 

We classify all the possible FCAs on a square and on a pentagonal graph, the latter representing the wrapped version of a FCA on $\mathbb Z$. We then study the solutions, providing a detailed analysis of the resulting dynamics.

Section \ref{sec:homoFCA} is devoted to a brief introduction to homogeneous FCAs, specifically focusing on the theoretical requirements adopted along the paper. We show how Group Theory is connected to the study of homogeneous FCAs. In Sec.~\ref{sec:UnivQW} we then show the first results concerning unitarity conditions for FCA. In Sec.~\ref{sec:casespropc} the two case studies are presented and analysed, obtaining a full classification of the possible unitary evolutions in these two cases. In Sec.~\ref{sec:matrixev} the matrix form of the evolution operators is analysed, and their phenomenological behaviour is analysed. 
In sec.~\ref{sec:conclu} we summarize our results and comment on future developments.

\section{Homogeneous FCAs} \label{sec:homoFCA}
\subsection{Theoretical requirements}
Let us consider a countable set $G$ of local Fermionic modes (LFMs) $\rA_g$ \cite{BravyiKitaev02,DarianoManessi14}, with $g\in G\subseteq\nats$, described by Fermionic operators $\psi_g$ that obey standard Canonical Anticommutation Relations (CARs)
\begin{align}
\{\psi_g,\psi_f\}=0,\quad\{\psi_g,\psi^\dag_f\}=\delta_{f,g}I.
\end{align} 
LFMs will be often referred to as memory cells throughout the paper. 

A FCA is a discrete-step evolution of the global system $\rA_G:=\bigotimes_{g\in G}\rA_g$ (the tensor product symbol simply denotes parallel composition in the sense of Refs.~\cite{Chiribella10,DarianoManessi14}) satisfying three requirements: Reversibility, homogeneity and locality. Reversibility corresponds to the mathematical requirement of unitarity. Homogeneity essentially is the requirement that, from the point of view of the evolution rule, there is no privileged site $\rA_I$ in the network $\rA_G$ of memory cells. On a fundamental ground, as the FCA is a candidate to represent a physical law, the homogeneity requirement is closely related to the usual physical requirement of homogeneity, i.e.~that there is no privileged point in space-time. Locality corresponds to the requirement that the update of a cell $\rA_i$ in a single step can be affected by finitely many other cells, called {\em neighbours}, whose number is {\em uniformly bounded}, i.e.~there exists a finite integer $k\in\nats$ such that for every site $\rA_i$ the number of cells $n_i$ in its neighbourhood is bounded as $n_i\leq k$. 

In most of the literature so far, {\em linear} FCAs have been considered, i.e.~FCAs expressed by a map $\tT$ on the fermionic algebra such that $\psi_{i,t+1}=\tT(\psi_{i,t})=\sum_{j=1}^{n_i}A_{ij}\psi_{j,t}$. The aim of the present work is then to expand the analysis, encompassing the non-linear case besides the linear one, which is exhaustively covered by the the theory of Quantum Walks. 

We now proceed expressing the above requirements in formal terms. Since the local Fermionic algebra is finitely generated, the evolution step is completely specified provided that the evolution of an arbitrary field operator $\psi_g$, with $g\in G$, is specified. The evolution map $\tT$ is non-linear if it has the following general form
	\begin{align}
	\label{eq:ev_gen}
\tT(\psi_{g,t})=\sum_{\bvec{s},\bvec t} T_{g;\bvec{s},\bvec t} \, (\psi^\dag_{j_1,t})^{s_1}(\psi_{j_1,t})^{t_1}\ldots(\psi_{j_k,t}^{\dagger})^{s_k}(\psi_{j_k,t})^{t_k},
	\end{align}
where $\bvec s,\bvec t\in\{0,1\}^{\times K_g}$, $\bvec j=(j_1,\ldots,j_k)\in G^{\times K_g}$ has a dependence on $g$ as $\bvec j(g)$, and $K_g$ is the size of the neighbourhood of $g$, namely the subset $N^-_g\subseteq G$ containing all the cells $f\in G$ whose corresponding operators $\psi_{f,t},\psi^\dag_{f,t}$ are involved in the expression of Eq.~\ref{eq:ev_gen}. We will often denote by 
\begin{align*}
\psi_{g,t+1}^{\prime}:=\tT(\psi_{g,t})
\end{align*}
the output produced by the update rule when applied to the operator $\psi_{g,t}$.

The locality requirement amounts to the constraint $K_g\leq k<\infty$. Every system located in $g$ then evolves into a function of the operators corresponding to a finite neighbourhood $N^-_g$, whose algebra is generated by a finite number of field operators. 

The second requirement that we express in mathematical terms is homogeneity. Classically, space-time is homogeneous if its points cannot be absolutely discriminated. In this FCA framework there is no pre-defined space-time structure to refer to in order to define homogeneity. Nevertheless, we can give a consistent definition based only on the elements introduced so far. Following Ref.~\cite{Perinotti17} for the definition of operational equivalence and absolute and relative discrimination, a rigorous statement of homogeneity is the following: \emph{the evolution rule allows for the discrimination of any two arbitrary systems $\mathrm{A}_{g_1}$ and $\mathrm{A}_{g_2}$, but only with respect to an arbitrary reference system $\mathrm{A}_{e}$ with $e \in G$}.

The above requirement largely simplifies the expression of the evolution rule of Eq.~\eqref{eq:ev_gen}. The first consequence is that $N^-_g$ has the same size $k$ for every $g\in G$. Moreover, homogeneity imposes that the coefficients $T_{g;\bvec{s},\bvec t}$ are the same at every $g\in G$, thus providing a natural way of establishing a bijective correspondence between any pair of neighbourhoods $N^-_g$ and $N^-_{g'}$ for $g,g'\in G$. This correspondence establishes an ordering $\bvec j(g)$ for every neighbourhood $N^-_g$, namely a correspondence with $S_+:=\{h_1,h_2,\ldots,h_k\}$. Eq.~\eqref{eq:ev_gen} thus becomes
	\begin{equation}
	\label{eq:ev_homog}
	\psi_{g,t+1}^{\prime} = \sum_{\bvec{s},\bvec t} T_{\bvec{s},\bvec t} (\psi^\dag_{j_1,t})^{s_1}(\psi_{j_1,t})^{t_1}\ldots(\psi_{j_k,t}^{\dagger})^{s_k}(\psi_{j_k,t})^{t_k},
	\end{equation}
where for the sake of brevity it is meant that $\bvec j=\bvec j(g)$.

\subsection{Neighbourhoods and Cayley graphs of groups}

We briefly review here the notion of a Cayley graph, and the way in which it is built from a homogeneous cellular automaton. Let $G$ denote the array of memory cells of a homogeneous FCA, and let $S_+$ be the set that is in correspondence with the neighbourhood $N^-_g$ for every $g\in G$. In particular, two elements $j_a\in N^-_{g_1}$ and $j_b\in N^-_{g_2}$ are in correspondence with the same $h_l\in S_+$ iff $j_a=j_l(g_1)$ and $j_b=j_l(g_2)$. One can then build a graph with vertex set $G$ and edge set $E\subseteq G\times G$, where $(f,g)\in E$ if $g\in N^-(f)$. If the graph is disconnected, it will consist of totally equivalent connected components, and then we will pick any connected component and restrict to the case where the graph is connected. If one goes into further detail, since every set $N^-_g$ is ordered by $S_+$, one can ``colour'' the edges by the colours $h_i\in S_+$, namely $(f,g)=h_l$ if $g=j_l(f)$. In this case we will use the shorthand $g=fh_l$, and equivalently $f=gh_l^{-1}$. Thanks to the homogeneity requirement, there are no equivalent colours (for a thorough proof of this fact, we refer to Ref.~\cite{Perinotti17}), and thus the choice of colouring is unique. If we identify a reference cell, and denote it by $e$, we will abbreviate $h_l:=eh_l=j_l(e)$, and then, recursively applying our notation, every $g\in G$ can be written as a word in the alphabet $S:=S_+\cup S_-$, with $S_-:=S_+^{-1}$. This construction defines a free group $F$ on the generators $S_+$.

\begin{figure}
		\centering
		\includegraphics[width=.49\columnwidth]{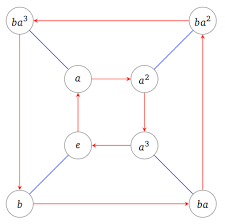}
		\includegraphics[width=.49\columnwidth]{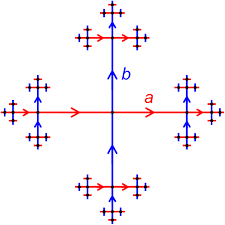}
		\caption{Two examples of Cayley graphs. On the left, the graph related to the dihedral group Dih4 on two generators $a$ and $b$ and on the right the graph related to the free group on two generators $a$ and $b$. \label{fig:cayley}}
	\end{figure}

One can now prove (see e.g.~Refs.~\cite{DAriano:2014aa,DAriano2016}) that, thanks to homogeneity, the closed paths on the graph correspond to the same sequences of colours, independently of what vertex one starts from. Thus, one can find a normal subgroup $R$ in $F$, corresponding to the normal closure of the set of words $w=e$. The quotient $F/S_+$ is a group, that we will denote by $G$. In particular, the graph that we constructed starting from the FCA corresponds to a Cayley graph $\Gamma(G,S_+)$. A Cayley graph is a graphical way of expressing a presentation $G::\langle S\mid R\rangle$ of the group $G$ in terms of a generating set $S_+$ and a set of relators $R$, whose normal closure corresponds to all the products of generators that amount to the identical element $e\in G$. The graph has $G$ as its vertex set, and $E=\{(g,gh)\mid h\in S_+\}$. It is easily proved that, for a given group $C$, neither the set $S_+$ nor the set $R$ are unique; on the other hand, any presentation completely specifies $C$. 
See Fig.~\ref{fig:cayley} for a couple of examples of Cayley graphs.

Following Refs.~\cite{DAriano2016,Perinotti17}, we can then rigorously associate the local modes of a homogeneous FCA with the nodes of the Cayley graph of some group $G$. Different FCAs, with different neighbourhood schemes, determine different Cayley graphs. 
\subsection{Wrapping lemma and CARs}
This section is devoted to the generalization of the result known as {\em wrapping lemma} (WL), stated in Ref.~\cite{Schumacher:2004aa} for a general QCA. WL states that \emph{the local transition rules $\tT_x$ of single memory cells $x\in G$ uniquely determine the global isomorphism $\tT$. Consequently, evolved algebras of cells whose neighbourhoods have no common sites commute elementwise}.
Eventually, by the homogeneity requirement, one has that $\tT_x$ is the same map for every cell $x$. 
A Fermionic version of the above lemma allows us to study a FCA on an infinite number of Local Fermionic Modes {\em via} a FCA on finitely many cells. We now give the Fermionic version of the lemma.

We identify the single site algebra $\tilde{\mathcal F}_1$ generated by the abstract fermionic operators $\psi,\psi^\dag$, with the subalgebra $\mathcal F_x$ of a given site $x$ by the embedding $\mathscr E_x:\psi\mapsto\psi_x$. More generally, the embedding acts on a local algebra $\mathcal F_n$ involving $n$ Local Fermionic Modes, as $\mathscr E_\Delta:\tilde{\mathcal F}_{|\Delta|}\to\mathcal F_\Delta$, where $\tilde{\mathcal F}_n$ is the algebra of a system made of $n$ fermionic modes, generated by $\psi_i,\psi_i^\dag$, $1\leq i\leq n$, while $\mathcal F_\Delta$ is the algebra generated by $\psi_x,\psi_x^\dag$ for $x\in\Delta$. The local rule $\tT_x$ maps $\mathcal F_x$ into $\mathcal F_{N^-_x}$, and more generally $\prod_{x\in\Delta}\tT_x$ maps $\mathcal F_\Delta$ into $\mathcal F_{N^-_\Delta}$, where $N^-_\Delta:=\bigcup_{x\in\Delta}N^-_x$. So, as $\tT$ is an algebra isomorphism, we have that 
\begin{equation}\label{eq:WRcond}
\tT\left( \prod_{x \in \Delta} \psi_x  \right) = \prod_{x \in \Delta} \tT_x(\psi_x)=\prod_{x \in \Delta} \tT_0(\psi_x),
\end{equation} 
where we used homogeneity to identify $\tT_x=\tT_0$, with $\tT_0=\mathscr E_{N^-_\Delta}\tilde\tT_0\mathscr E^{-1}_\Delta:\mathcal F_\Delta\to\mathcal F_{N^-_\Delta}$ is the embedding of $\tilde\tT_0:\tilde{\mathcal F}_1\to\tilde{\mathcal F}_{|N^-_x|}$.
Since on the left hand side we have an expression generated by the images of the generators $\psi_x$ of the Fermionic algebra, which are isometrically embedded in a sub-algebra of the operators on the neighbourhood $N^-_x$, we can exploit the finiteness of the latter to simplify the unitarity conditions for the FCA. A necessary condition for unitarity is the preservation of the canonical anti-commutation relations, i.e. 
\begin{align*}
\{\psi'_x,\psi'_y\}=0,\quad\{\psi'_x,(\psi'_y)^\dag\}=\delta_{x,y}I.
\end{align*}
For finite graphs the above condition is necessary and sufficient for unitarity, due to the uniqueness of the representation of the Fermionic algebra up to unitary transformations \cite{Nielsen05}. 

The anti-commutativity condition is not trivial because of the possible overlappings among different neighbourhoods. 
Since only a small portion of the lattice is needed to verify the reversibility conditions for the global update rule $\tT$ in terms of the local update rule $\tT_0$ of Eq.~\eqref{eq:WRcond}, we consider a new lattice characterised by the same neighbourhood structure as the original one, plus some {\em periodic boundary conditions} ({\em p.b.c}). A precise way to introduce {\em p.b.c} is by taking the finite quotient of the group $G$ by a normal subgroup $\Pi$: given a homomorphism $\phi: \Gamma \mapsto \Xi$, with $\Xi$ a finite group, we can identify $\Pi\subseteq\Gamma$ as $\Pi = \mathrm{Ker}\phi$. We define the {\em p.b.c} so that the memory cells of the periodic system are those differing by an element $\pi \in \Pi$. Consequently, the set of cells we are interested in are in the finite quotient $\Xi = \Gamma/ \Pi$. Since we work now with a finite quotient group, we are allowed to verify unitarity as a condition of CARs preservation \cite{Nielsen05}. However, we must take care that the quotient does not introduce extra conditions for the local rule. We say that a neighbourhood $N^-_g$ is {\em regular} for the periodic structure given by $\Xi$ if the equations for the anti-commutation of the evolved field operators are the same for the local rule on $\Xi$ and that on $G$. 
 

A neighbourhood $N^-$ is $regular$ for a normal subgroup $\Pi$ if $N_{[x]}^-\cap N^-_{[x]h_1h_2^{-1}}=[{N_{x}^-\cap N^-_{xh_1h_2^{-1}}}]$, where $[x]\in\Xi$ is the equivalence class of $x\in\Gamma$. We have thus translated in terms of Fermionic algebras and the construction needed to prove the so-called {\em Wrapping Lemma} of Ref.~\cite{Schumacher:2004aa}. The different $p.\ b.\ c.$ group construction and the request of {\em anti-commutation} rather than commutation do not undermine the argument given in Ref.~\cite{Schumacher:2004aa} that can be now trivially and properly adapted to obtain the following Fermionic version of the WL: \emph{The FCA transition rules on a finite lattice $\Xi=G/\Pi$ with respect to a regular neighbourhood scheme $N^-_g$ are in one-to-one correspondence with the transition rules for FCAs on $G$ with the same neighbourhood scheme.}

\section{Universality of Quantum Walk conditions}\label{sec:UnivQW}
In this section we present the first result of the paper, that regards the linear terms in the expression of the evolved field operators in Eq.~\eqref{eq:ev_homog}. Indeed, we now prove that under general circumstances the coefficients for the linear terms must always satisfy unitarity conditions of Quantum Walks (QWs). 

To show this result we start from Eq.~\eqref{eq:ev_homog}. Imposing unitarity of the evolution through the preservation of CARs requires calculating anti-commutation parentheses between different odd field polynomials, as clear from Eq.~\eqref{eq:ev_homog}. 
We now show that no anti-commutation but those involving two conjugate field operators can produce an identity operator $\mathbb{I}$. Indeed, the operator $\mathbb{I}$ appears as the result of an anti-commutation $\{\psi_x, \psi_x^{\dagger} \}= \mathbb{I}$, or in the normal ordering of an anti-ordered operator, i.e.~$\psi_x\psi_x^{\dagger}= \mathbb{I} -\psi_x^{\dagger}\psi_x$.

While the first instance is peculiar of QWs, and shows up in the non-linear case too, the second one is peculiar of the non-linear case. Nevertheless, it is never the case that a term proportional to the identity operator is produced as in the second instance, namely by reordering of an anti-ordered term. 

To show this, we divide the set of possible monomials in the right hand side of Eq.~\ref{eq:ev_homog} in two subsets: Monomials involving one or more number operators (of the form $\psi^\dag_y\psi_y$, with $y\in N^-_g$) and monomials that do not. For every number operator $\psi^\dag_y\psi_y$ involved in a monomial, the result of the anti-commutation with a different monomial will involve either $\psi^\dag_y\psi_y$ itself, or a field operator $\psi^\dag_y$, or $\psi_y$. Indeed, denoting by $A$ and $B$ two nonlinear operators whose expression does not include $\psi_x$ or $\psi^\dag_x$, we have:
\begin{align*}
&\{A\psi_x, B \psi_x^{\dagger}\psi_x\}=s_1 AB \psi_x^\dag,\\
&\{A\psi^\dag_x, B \psi_x^{\dagger}\psi_x\}=s_2 BA \psi_x,\\
&\{A\psi_x^{\dagger}\psi_x, B \psi_x^{\dagger}\psi_x\}=\{A,B\} \psi_x^{\dagger}\psi_x,
\\&\{A, B \psi_x^{\dagger}\psi_x\}=\{A,B\} \psi_x^\dag\psi_x,
\end{align*}
where $s_1$ and $s_2$ are signs depending on the number of field operators in the nonlinear operator $A$ and $B$. Therefore, anticommutations having a monomial of the first kind as an argument cannot give the identity operator as a result. 

We now introduce the expression ``$\xi$-like terms", to denote those monomials whose expression does not involve any number operator. For example, $\psi_x\psi_y\psi^\dag_z$ is a $\xi$-like term, while $\psi^\dag_x\psi_x\psi_y$ is not. The above terminology is due to the notation that we use in the following. Every field operator involved in a $\xi$-like term refers to a different site of the neighbouhood. We can therefore say that these terms are in the form of $A\psi_x$ or $B\psi_x^{\dag}$, where $A, B$ are nonlinear operators that do not involve $\psi_x$ nor $\psi^\dag_x$. Let us now consider
\begin{equation}
\begin{split}
\{A\psi_x^{\dagger}, B \psi_x\}&=A\psi_x^{\dagger}B\psi_x + B\psi_xA\psi_x^{\dag} \\ &= s_1AB\psi_x^{\dag}\psi_x + s_2BA(\mathbb I-\psi_x^{\dag}\psi_x), 
\end{split}
\end{equation}
where $s_1, s_2$ are signs depending on how many field operators are in $A, B$. So
\begin{equation}\label{eq:xicond}
\{A\psi_x^{\dagger}, B \psi_x\} = (s_1AB - s_2BA)\psi_x^{\dag}\psi_x + s_2BA.
\end{equation}
It is easy to realize that  the right-hand side of Eq.~\ref{eq:xicond} cannot equal the identity operator unless $A$ and $B$ are both identity.

We have then identified the first general condition of unitarity for a nonlinear evolution of a FCA: The normalization conditions for Quantum Walks holds for the coefficients of degree-one monomials of the evolved field operators under any FCA.
As a consequence, every non-linear automaton can be thought of as a linear automaton (free evolution of Fermionic excitations, or ``particles" for short) augmented with a non-linear interacting term between particles.

\section{Case studies}\label{sec:casespropc}
Here we report the calculation of the unitary conditions, along with their solutions, for two special cases of scalar FCAs. These are FCAs whose Cayley graphs are finite, and correspond to the following presentations $\left\langle a, b | a^2, b^2, (ab)^2 \right\rangle$ and $\left\langle a | a^5 \right\rangle$, of the groups $\mathbb Z_2\times\mathbb Z_2$ and $\mathbb Z_5$, respectively. 
\begin{figure}
	\centering
	\includegraphics[width=.49\columnwidth]{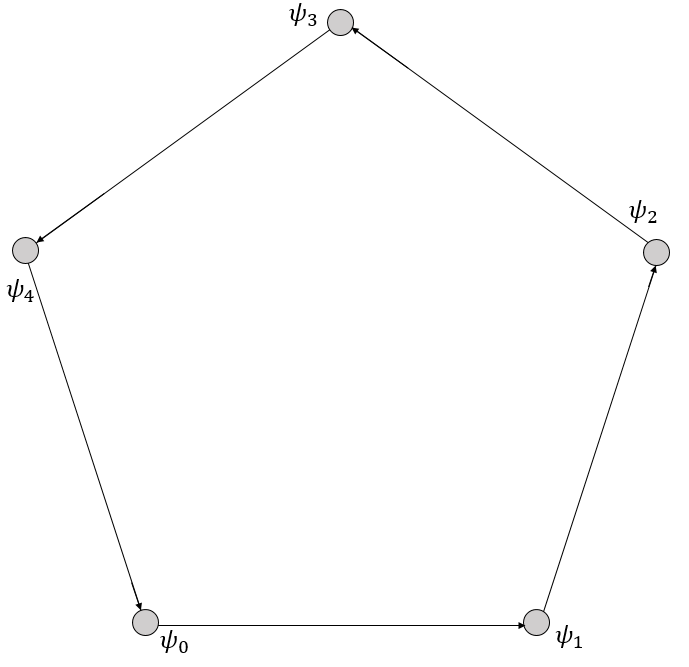}
	\includegraphics[width=.49\columnwidth]{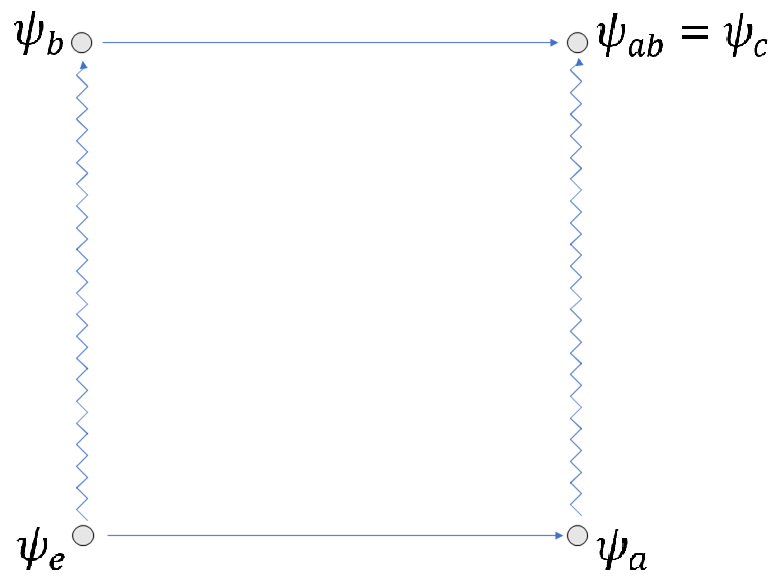}
	\caption{Cayley graph related to the presentations of the groups analysed in the present case studies. On the left the graph related to $\left\langle a | a^5 \right\rangle$ and on the right the graph related to $\left\langle a, b | a^2, b^2, (ab)^2 \right\rangle$.}
	\label{fig:casestudies}
\end{figure}
As specified in Ref.~\cite{Nielsen05}, preservation of the CARs is equivalent to unitarity of the evolution for FCAs on finite groups. We will then express the unitarity conditions as a set of second order equations produced by constraining the anti-commutation of polynomials expressing the evolved field operators.

First of all we observe that, following Ref.~\cite{Ostlund91}, monomials of even degree are excluded in the expression of Eq.~\eqref{eq:ev_homog}. Moreover, every term exhibiting the same field operator more than once is inevitably null because of CARs. In particular, as a consequence of the last observation, it is not restrictive to impose that the strings $\bvec j(g)\in G^{\times k}$ in Eq.~\eqref{eq:ev_homog} cannot have $j_i(g)=j_l(g)$ for $i\neq l$. Given these constraints, one can easily verify that we can have terms of degree one, three and five in Eq.~\eqref{eq:ev_homog} in the cases under consideration (see Fig.~\ref{fig:casestudies}). 

\begin{figure}
		\centering
		\includegraphics[width=.75\columnwidth]{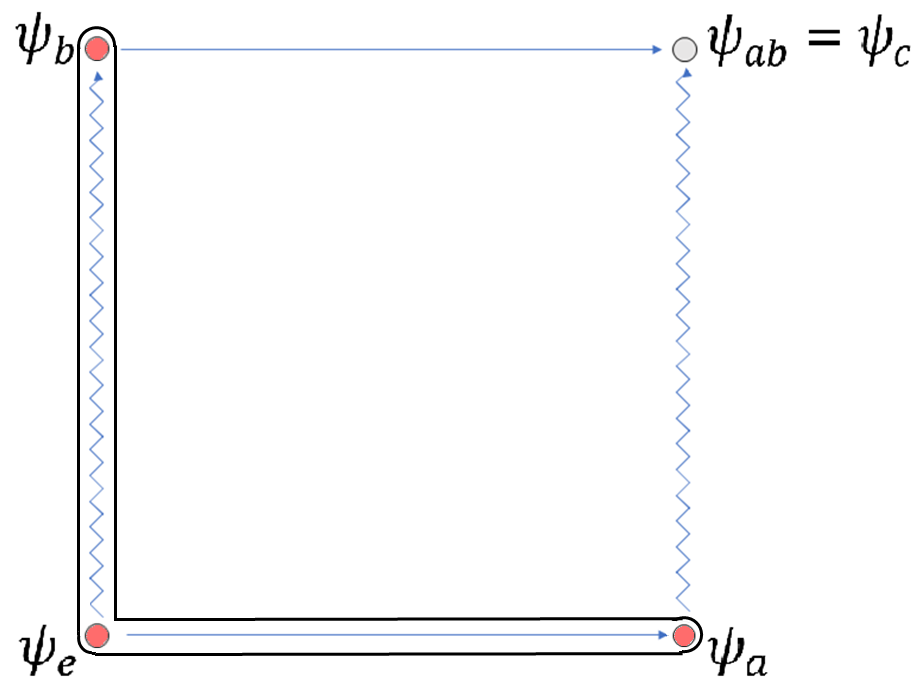}
		\caption{Graphical representation of a non-linearly spread operator on the Cayley graph of $\mathbb Z_2\times \mathbb Z_2$, the starting site being $e$, with black lines edging the neighbourhood $N^-_e$ where the operator is spread.}
		\label{fig:spredop}
	\end{figure}

\begin{figure}\label{fig:steps}
	\centering
	{\includegraphics[width=.49\columnwidth]{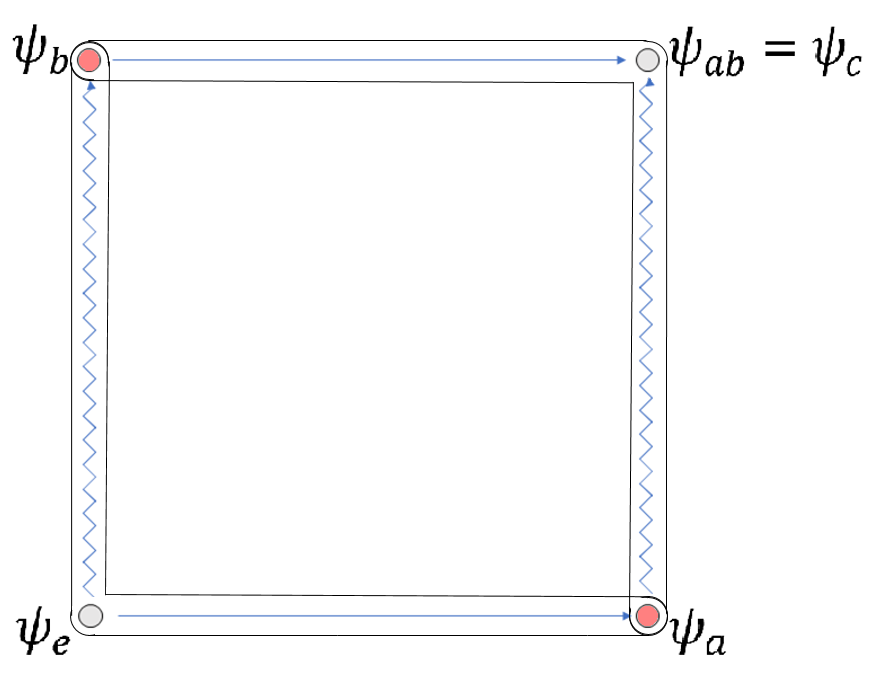}} 
	{\includegraphics[width=.495\columnwidth]{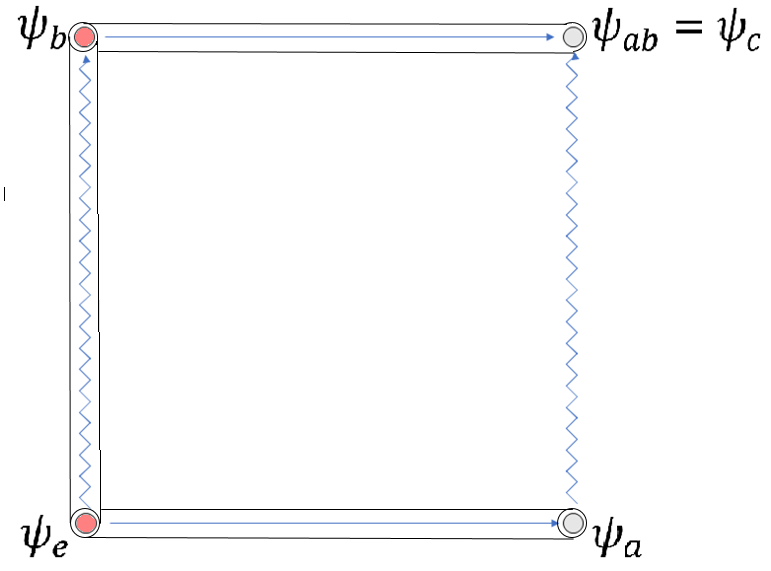}} \\
	{\includegraphics[width=.49\columnwidth]{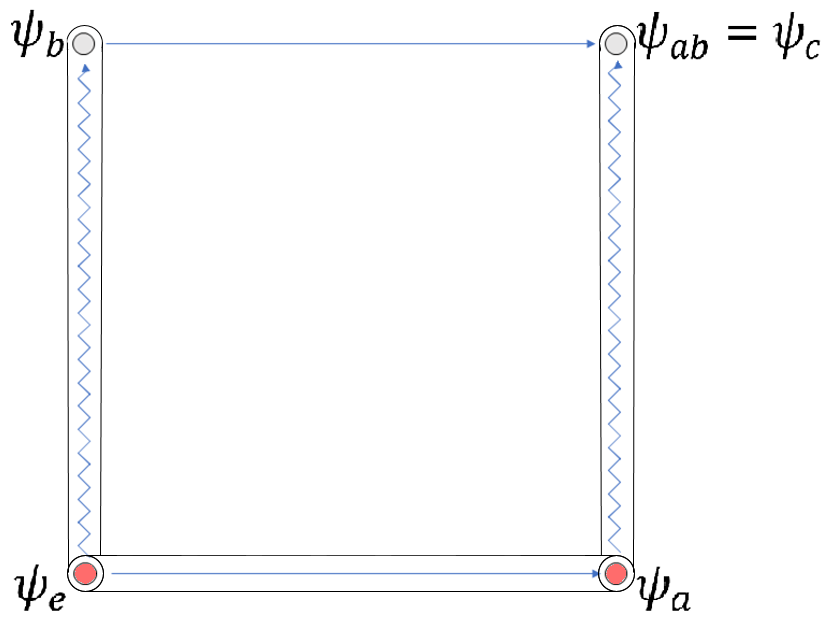}} 
	{\includegraphics[width=.495\columnwidth]{Cayley_spread}} 
	\caption{Different overlapping between spread operators. Sites shared by the two neighbourhoods are highlighted in red.}
\end{figure}

While linear terms describe {\em transitions} of information toward neighbouring sites (like transition matrices for Quantum Walks), non-linear terms describe spreading of information, in the sense that information flows from one site toward multiple sites at once. For this reason, we call the non-linear terms {\em spread terms}. A graphical illustration of spreading on the Cayley graph of $\mathbb Z_2\times\mathbb Z_2$ is illustrated in Fig.~\ref{fig:spredop}, where the neighbourhood $N^-_e$ of the site $e$ is highlighted.
In both cases, the calculation is divided in three steps, in order to ease the analysis. Since we calculate anti-commutation parentheses among pairs of spread terms, we start the calculation from the least overlapped pairs of neighbourhoods, and then we proceeded with increasingly overlapping pairs, ending with the anti-commutation of evolved field operators with themselves. Fig.~\ref{fig:steps} shows the steps of this procedure. 
\subsection{Group $\mathbb Z_2\times\mathbb Z_2$}\label{sec:ab}
In this section we present the analysis of the first case study. In this case, we require the update rule to be number-preserving, i.e.~to preserve the total number of Fermionic excitations. This implies a further constraint on the expression of the evolved operators in Eq.~\eqref{eq:ev_homog}. We refer to Fig.~\ref{fig:casestudies} (right) for a graphical representation of the Cayley graph associated to the group presentation. A representation $\tZ$ of the the group generators acts on the Fermionic algebra as follows: $\tZ_g(\psi_f):=\psi_{gf}$. We have four different field operators, $\psi_e, \psi_a, \psi_b$ and $ \psi_{ab}$, obtained by the action of the representation $\tZ$ on $\psi_e$. The sites $a$ and $b$ are the neighbours of the site $e$. We use $c$ as a shorthand for ${ab}$, and we order the basis of the Fermionic algebra as follows: $(\psi_a,\psi_b,\psi_e,\psi_c)$.

We start imposing the conditions $\{\psi_e^{\prime}, \psi_c^{\prime}\}=0$, $\{\psi_e^{\prime},(\psi_c^{\prime})^{\dagger}\}=0$. From Eq.~\eqref{eq:ev_homog} and the preliminary considerations we can write the explicit form of $\psi_e^{\prime}$ and $\psi_c^{\prime}$:
\begin{equation}
	\label{psi_e}
	\begin{split}
	\psi_e^{\prime} &=\alpha^e_a \psi_a + \alpha^e_b \psi_b + \alpha^e_e \psi_e + \beta^e_{ae} \psi_a^{\dagger}\psi_a\psi_e + \\&\beta^e_{ea} \psi_a\psi_e^{\dagger}\psi_e + \beta^e_{ba} \psi_a\psi_b^{\dagger}\psi_b + \beta^e_{ab} \psi_a^{\dagger}\psi_a\psi_b +\\& \beta^e_{eb} \psi_b\psi_e^{\dagger}\psi_e + \beta^e_{be} \psi_b^{\dagger}\psi_b\psi_e + \xi^e_{abe}\psi_a^{\dagger}\psi_b\psi_e + \\&\xi^e_{bea}\psi_a\psi_b^{\dagger}\psi_e + \xi^e_{eab}\psi_a\psi_b\psi_e^{\dagger} + \gamma^e_{abe}\psi_a^{\dagger}\psi_a\psi_b^{\dagger}\psi_b\psi_e + \\&\gamma^e_{bea}\psi_a\psi_b^{\dagger}\psi_b\psi_e^{\dagger}\psi_e + \gamma^e_{eab}\psi_a^{\dagger}\psi_a\psi_b\psi_e^{\dagger}\psi_e
	\end{split}
	\end{equation}
and
	\begin{equation}
	\label{psi_c}
	\begin{split}
	\psi_c^{\prime} &=\alpha^c_a \psi_b + \alpha^c_b \psi_a + \alpha^c_c \psi_c + \beta^c_{ac} \psi_b^{\dagger}\psi_b\psi_c +\\& \beta^c_{ca} \psi_b\psi_c^{\dagger}\psi_c + \beta^c_{ba} \psi_b\psi_a^{\dagger}\psi_a + \beta^c_{ab} \psi_b^{\dagger}\psi_b\psi_a +\\& \beta^c_{cb} \psi_a\psi_c^{\dagger}\psi_c + \beta^c_{bc} \psi_a^{\dagger}\psi_a\psi_c + \xi^c_{abc}\psi_b^{\dagger}\psi_a\psi_c + \\&\xi^c_{bca}\psi_b\psi_a^{\dagger}\psi_c + \xi^c_{cab}\psi_b\psi_a\psi_c^{\dagger} + \gamma^c_{abc}\psi_b^{\dagger}\psi_b\psi_a^{\dagger}\psi_a\psi_c +\\& \gamma^c_{bca}\psi_b\psi_a^{\dagger}\psi_a\psi_c^{\dagger}\psi_c + \gamma^c_{cab}\psi_b^{\dagger}\psi_b\psi_a\psi_c^{\dagger}\psi_c.
	\end{split}
	\end{equation}
Notice the presence of trilinear $\xi$-like terms, whose coefficients are denoted by the symbol $\xi$. We highlighted the role of this kind of terms in the generalization of unitarity conditions in Sec.~\ref{sec:UnivQW}. Coefficients in Eq.~\eqref{psi_e} and Eq.~\eqref{psi_c} have lower and upper indices: Since every term in Eq.~\eqref{psi_e} and Eq.~\eqref{psi_c} is a spread operator, the lower indices denote the sites where the term is spread, while the upper index denotes the starting site, namely the site $g$ in Eq.~\eqref{eq:ev_homog}. 

Homogeneity provides a relation between coefficients in Eq.~\eqref{psi_e} and those in Eq.~\eqref{psi_c}. Indeed, the lower indices of the coefficients refer to subsets of the neighbourhood. Homogeneity allows us to identify coefficients of homologous subsets of different neighbourhoods. For example, we will have $\beta^e_{ae}=\beta^c_{bc}$.
We then have
	\begin{align*}
	\alpha^e_a &=\alpha^c_b & \beta^e_{ea} &=\beta^c_{cb} & \beta^e_{ae} &=\beta^c_{bc}\\ 
	\alpha^e_b &=\alpha^c_a & \beta^e_{eb} &=\beta^c_{ca} & \beta^e_{be} &=\beta^c_{ac}\\ 
	\alpha^e_e &=\alpha^c_c & \beta^e_{ab} &=\beta^c_{ba} & \beta^e_{ba} &=\beta^c_{ab} 
	\end{align*}
	\begin{align*}
	\xi^e_{abe} &= \xi^c_{bac} = -\xi^c_{bca} & \gamma^e_{bea} &= \gamma^c_{acb}=\gamma^c_{cab}\\
	\xi^e_{bea} &= \xi^c_{acb} = -\xi^c_{abc} & \gamma^e_{abe} &= \gamma^c_{bac}=\gamma^c_{abc}\\
	\xi^e_{eab} &= \xi^c_{cba} = -\xi^c_{cab} & \gamma^e_{eab} &= \gamma^c_{cba}=\gamma^c_{bca}.\\
	\end{align*}
Imposing the condition $\{\psi_e^{\prime}, (\psi^\dag_c)^{\prime}\}=0$, we obtain that $\xi^e_{eab}$ must be null and the other two $\xi$ coefficients must be opposite. We can thus appreciate the convenience of the present procedure, as in the following we can simplify the expressions of the evolved field operators setting $\xi^e_{eab}=0$, and referring to the other two $\xi$ coefficients as $\pm \xi^e$.
	
The second set of anti-commutators is $\{\psi_e^{\prime}, \psi_a^{\prime}\}=0$, $\{\psi_e^{\prime},(\psi_a^{\prime})^{\dagger}\}=0$  and $\{\psi_e^{\prime}, \psi_b^{\prime}\}=0$, $\{\psi_e^{\prime},(\psi_b^{\prime})^{\dagger}\}=0$. We do not report the explicit forms of $\psi'_a$ and $\psi'_b$, since they are obtained applying the same combinations of generators as in Eqs.~\eqref{psi_e} and~\eqref{psi_c}. Homogeneity allows us to identify the following coefficients of $\psi'_e$ and $\psi'_a$:
	\begin{align*}
	\alpha^e_a &=\alpha^a_e & \beta^e_{ea} &=\beta^a_{ae} & \beta^e_{ae} &=\beta^a_{ea} & \gamma^e_{eab} &= \gamma^a_{aec}\\
	\alpha^e_b &=\alpha^a_c & \beta^e_{eb} &=\beta^a_{ac} & \beta^e_{be} &=\beta^a_{ca} & \gamma^e_{bea} &= \gamma^a_{cae}\\
	\alpha^e_e &=\alpha^a_a & \beta^e_{ab} &=\beta^a_{ec} & \beta^e_{ba} &=\beta^a_{ce} & \gamma^e_{abe} &= \gamma^a_{eca}.
	\end{align*}
In the same way, we can identify the following coefficients of $\psi'_e$ and $\psi'_b$:
	\begin{align*}
	\alpha^e_a &=\alpha^b_c & \beta^e_{ea} &=\beta^b_{bc} & \beta^e_{ae} &=\beta^b_{cb} & \gamma^e_{eab} &= \gamma^b_{bce}\\
	\alpha^e_b &=\alpha^b_e & \beta^e_{eb} &=\beta^b_{be} & \beta^e_{be} &=\beta^b_{eb} & \gamma^e_{bea} &= \gamma^b_{ebc} \\
	\alpha^e_e &=\alpha^b_b & \beta^e_{ab} &=\beta^b_{ce} & \beta^e_{ba} &=\beta^b_{ec} & \gamma^e_{abe} &= \gamma^b_{ceb}.
	\end{align*}
Recollecting the independent equations obtained by imposing the six anti-commutations, we obtain the following second-degree system of equations
	\begin{align}\label{sis:a-b-ab}
	&\alpha_x\beta_{yz}=0\nonumber\\
	& \beta_{xy}\beta_{yz}=0 \nonumber\\ 
	&\alpha_x\beta_{yz}^*+\alpha_z\beta_{yx}^*=0 \nonumber\\
	&\beta_{xy}\beta_{xz}^*+\beta_{xz}\beta_{xy}^*=0 \nonumber\\
	&\alpha_x\alpha_y^*+\alpha_y\alpha_x^*=0 \nonumber\\
	& \alpha_x\beta_{xy}-\alpha_y\beta_{yx}=0\nonumber\\
	&\alpha_z\gamma_{xyz}-\alpha_y\gamma_{zxy}+\beta_{xy}\beta_{zy}-\beta_{xz}\beta_{yz}=0 \nonumber\\
	& \alpha_x\beta_{yx}^*+\alpha_x^*\beta_{yx}+\beta_{yx}\beta_{yx}^*=0 \nonumber\\
	&\alpha_z\beta_{zy}^*+\beta_{yz}\alpha_y^*+\beta_{yz}\beta_{zy}^*=0 \\
	& \beta_{xy}\gamma_{ijk}\delta_{yj}+\beta_{zw}\gamma_{kij}\delta_{kw}+\gamma_{ijk}\gamma_{kij}=0\nonumber\\
	&\alpha_x\gamma_{xyz}+\beta_{yx}\beta_{yz}+\beta_{yx}\gamma_{xyz}=0 \nonumber\\
	& \alpha_x\gamma_{zxy}+\beta_{zx}\beta_{zy}+\beta_{zx}\gamma_{zxy}=0\nonumber\\
	&\alpha_x\gamma_{ijk}^*+\beta_{yx}\gamma_{ijk}^*+\delta_{kx}\beta_{yx}\beta_{zx}^*+\delta_{ix}\beta_{yx}\beta_{jx}^*\nonumber\\
	& +\beta_{yx}\beta_{ik}^*\delta_{jx}=0\nonumber\\
	&\beta_{xy}\gamma_{ijk}^*+\gamma_{zxy}(\beta_{xy}^*\delta_{ky}+\beta_{zx}^*\delta_{kx}+\beta_{xz}^*\delta_{kz})\nonumber\\
	& +\gamma_{zxy}\gamma_{ijk}^*=0\nonumber
	\end{align}
	where $x, y, z$ can be any of the neighbouring sites, appropriately ordered. The above expressions are clearly cyclic in the variables $x,y,z$.
	
We remark that, besides the system of Eq.~\eqref{sis:a-b-ab}, one obtains independent equations that we did not report, which impose $\xi^e = \xi^a=\xi^b$ and $\xi^e=0$. As a consequence, no $\xi$ coefficients appear in the expression of $\psi'_g$ under the unitary evolution of a homogeneous FCA on the Cayley graph corresponding to $\left\langle a, b | a^2, b^2, (ab)^2 \right\rangle$.

Having calculated the unitarity conditions, expressed by the system in Eq.~\eqref{sis:a-b-ab}, we can proceed to their solution. Expressing all the coefficients in polar form, we have  
	\begin{align*}
&\alpha_x = |\alpha_x| e^{i \theta_{x}}, &&\beta_{xy} = |\beta_{xy}| e^{i \theta_{xy}}. 
\end{align*}

We can identify three families of solutions, grouped by the subset of non-null linear coefficients $\alpha_x$, with $x=a, b, e$. We report here the families of solutions.
	
\begin{enumerate}
\item $\alpha_x\neq0 \quad\forall \ x=a, b, e$: No solutions are admitted.
\item $\alpha_x\neq0$ for only one value of $x=a,b,e$: The non-null coefficients are $\alpha_x,\beta_{ix},\beta_{jx},\gamma_{ijx} \neq 0$ for $i,j \in \{a,b,e\}\setminus\{x\}$. The conditions on the coefficients are
\begin{subequations}
	\begin{equation}
	\alpha_x\alpha_x^* = 1,
	\end{equation}
	\begin{equation}
	\gamma_{ijx} = \beta_{ix}\beta_{jx},
	\end{equation}
	\begin{equation}\label{eq:condBeta1}
	2\cos(\varphi_{ix}) = - |\beta_{ix}|,
	\end{equation}
	\begin{equation}\label{eq:condtheta1}
	\varphi_{ix}=\varphi_{jx},
	\end{equation}
\end{subequations}
and consequently
\begin{equation}\label{eq:condBeta1def}
\beta_{ix}=\beta_{jx}=:\beta = |\beta| e^{\varphi}.
\end{equation}
\item $\alpha_x, \alpha_y\neq0$ for two values $x\neq y$ and $x,y=a,b,e$: The non-null coefficients are $\alpha_x,\alpha_y,\beta_{yx},\beta_{xy}\neq0$. The conditions on the coefficients are
\begin{subequations}
	\begin{equation}
	\alpha_x \alpha_x^* + \alpha_y \alpha_y^* = 1
	\end{equation}
	\begin{equation}
	2|\alpha_i|\cos(\theta_i - \varphi_{ji}) = - |\beta_{ji}|
	\end{equation}
	\begin{equation}
	\theta_x - \varphi_{yx} = \theta_y - \varphi_{xy}.
	\end{equation}
\end{subequations}
\end{enumerate}
In Tab.~\ref{tab:1} we report a synoptic summary of the above families of solutions.
	\begin{table}[h]
		\begin{tabular*}{\columnwidth}{@{\extracolsep{\fill} }lll}
			\hline 
			$\alpha_x\neq0$ & $\alpha_x\neq0$ & $\alpha_x\neq0, \ \alpha_y\neq0 $ \\
			\ $x=a, b, e$ &\ $x=a, b, e$ &\ $x,y=a, b, e$ \\ 
			\hline 
			& $\alpha_x,\beta_{ix},\beta_{jx},\gamma_{ijx} \neq 0$  & $\alpha_x,\alpha_y,\beta_{yx},\beta_{xy}\neq0$  \\ 
			& $\alpha_x\alpha_x^* = 1$; & $\alpha_x \alpha_x^* + \alpha_y \alpha_y^* = 1$ \\ 
			\rule[-8 pt]{0 pt}{0 pt}
			\centering No FCA&$\gamma_{ijx} = \beta_{ix}\beta_{jx}$  &  \\
			\rule[-8 pt]{0 pt}{0 pt}
			&$2\cos(\varphi_{ix}) = - |\beta_{ix}|$  &$2|\alpha_i|\cos(\theta_i - \varphi_{ji}) =- |\beta_{ji}|$ \\
			& $\varphi_{ix}=\varphi_{jx}$ & $\theta_x - \varphi_{yx} = \theta_y - \varphi_{xy}$\\ 
			\hline 
		\end{tabular*}
		\caption{Family solution for $\left\langle a, b | a^2, b^2, (ab)^2 \right\rangle$ case. For the second family, we assume without loss of generality $\theta_x=0$. Indexes $i, j \in \{a, b, e\}$ always respect the normal order.}
		\label{tab:1}
	\end{table}

We remark that we cannot have a non-linear FCA with a trivial linear sector, in agreement with the results of Sec.~\ref{sec:UnivQW}, where we pointed out the universality of the Quantum Walk conditions on the linear terms. 

The second and third families of solutions show an evident symmetry: Both the non null $\beta$ coefficients satisfy the same relation between modulus and argument, and the corresponding relations for the two families are manifestly similar.
	
Moreover, in both the non trivial families of solutions we notice a spontaneous degree of isotropy in the information flow ruled by the FCA: In a sense, information flows symmetrically along the two directions on the graph, corresponding to the two different generators of the group. This spontaneous isotropy is confirmed in the second family of solutions by the condition \eqref{eq:condBeta1def}.
	
Finally, we notice that in the third family of solutions there is a natural limit to the non-linearity. Indeed, no pentalinear terms are non null.
	
We can greatly simplify the system \ref{sis:a-b-ab} by imposing perfect isotropy, i.e.~identifying the coefficients corresponding to permutation of the indices $a$ and $b$. In this case, one can find that every non-linear coefficient must be null, and the FCA reduces to a Quantum Walk.
\subsection{Group $\mathbb Z_5$}\label{sec:a}
We present here the results concerning the group $\left\langle a | a^5 \right\rangle$. 
	\begin{figure}
		\centering
		\includegraphics[width=.63 \columnwidth]{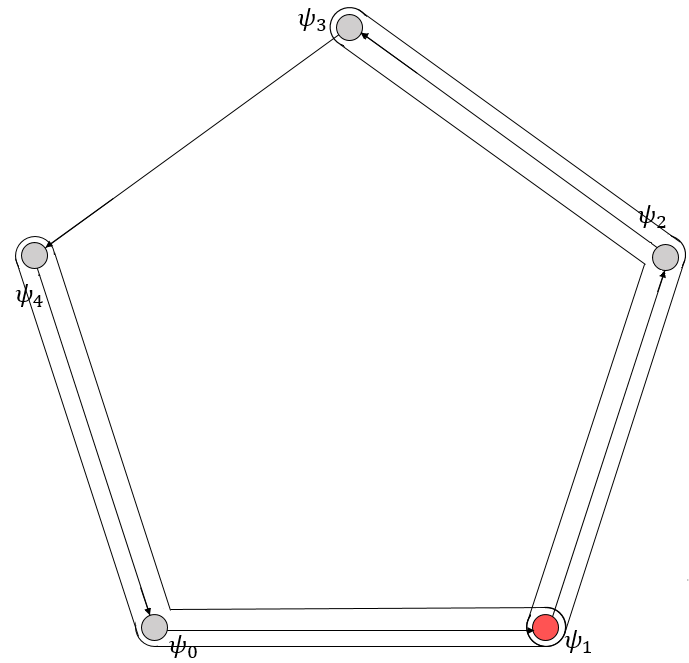}
		\caption{Graphic representation of the minimal overlapping between spread operators involved in $\{\psi_0^{\prime},\psi_2^{\prime}\}$ and $\{\psi_0^{\prime},(\psi_2^{\prime})^{\dagger}\}$.}
		\label{fig:overlapa5}
	\end{figure}
In this case we only require the update rule to preserve the CARs, so we allow for evolutions that are generally not number-preserving. In Fig.~\ref{fig:casestudies} (left) is the Cayley graph associated with the presentation $\langle a| a^5\rangle$ of the group $\mathbb Z_5$: there are five different fields $\psi_0, \psi_1, \psi_2, \psi_3, \psi_4$. Also in this case, a representation $\tZ$ of the the group acts on the Fermionic algebra as: $\tZ_g(\psi_f):=\psi_{gf}$, so that $\psi_k=\tZ_k(\psi_0)$, for $k=0,\ldots,4$. The neighbours of site $0$ are sites $1$ and $4$, and we order the basis of the Fermionic algebra as follows: $(\psi_1,\psi_0,\psi_4,\psi_3,\psi_2)$.
	
Paradigmatically, the unitarity conditions for this group are valid for every monogenerated group apart from $\mathbb Z_2$, $\mathbb Z_3$ and $\mathbb Z_4$. This is easily understood focusing on the graph in Fig.~\ref{fig:casestudies} (left). Suppose that the graph has its shortest relator longer than four. All the anti-commutation conditions that one has to impose, besides those of analyzed for $\mathbb Z_5$, are trivial, as third or further neighbourhoods are not overlapping, and the corresponding terms produced by the evolution are thus straightforwardly anti-commuting.

Let us now write down the expression for the evolved operator starting from site $0$:
	 \begin{equation}
	 \label{psi_0}
	 \begin{split}
	 \psi_0^{\prime}=& \alpha_0^{0}\psi_0 + \alpha_1^{0}\psi_1 + \alpha_4^{0}\psi_4 + \gamma_0^{0}\psi_0^{\dagger} +\\& \gamma_1^{0}\psi_1^{\dagger} + \gamma_4^{0}\psi_4^{\dagger} + \beta_{01}^{0}\psi_0^{\dagger}\psi_0\psi_1 +\\& \beta_{10}^{0}\psi_1^{\dagger}\psi_1\psi_0 + \beta_{04}^{0}\psi_0^{\dagger}\psi_0\psi_4 + \beta_{40}^{0}\psi_0\psi_4^{\dagger}\psi_4 +\\&
	 \beta_{14}^{0}\psi_1^{\dagger}\psi_1\psi_4 + \beta_{41}^{0}\psi_1\psi_4^{\dagger}\psi_4 + \xi_{104}^{0}\psi_1^{\dagger}\psi_0\psi_4 +\\& \xi_{041}^{0}\psi_1\psi_0^{\dagger}\psi_4 + \xi_{410}^{0}\psi_1\psi_0\psi_4^{\dagger} + \eta_{01}^{0}\psi_1^{\dagger}\psi_0^{\dagger}\psi_0 +\\& \eta_{10}^{0}\psi_1^{\dagger}\psi_1\psi_0^{\dagger} + \eta_{04}^{0}\psi_0^{\dagger}\psi_0\psi_4^{\dagger} + \eta_{40}^{0}\psi_0^{\dagger}\psi_4^{\dagger}\psi_4 +\\& \eta_{14}^{0}\psi_1^{\dagger}\psi_1\psi_4^{\dagger} + \eta_{41}^{0}\psi_1^{\dagger}\psi_4^{\dagger}\psi_4 + \chi_{104}^{0}\psi_1^{\dagger}\psi_0^{\dagger}\psi_4 + \\&\chi_{041}^{0}\psi_1\psi_0^{\dagger}\psi_4^{\dagger} + \chi_{410}^{0}\psi_1^{\dagger}\psi_0\psi_4^{\dagger} + \theta^{0}\psi_1\psi_0\psi_4 +\\& \bar{\theta}^{0}\psi_1^{\dagger}\psi_0^{\dagger}\psi_4^{\dagger} + \mu^0_{410}\psi_1^{\dagger}\psi_1\psi_4^{\dagger}\psi_4\psi_0 +\\& \mu^0_{041}\psi_1\psi_4^{\dagger}\psi_4\psi_0^{\dagger}\psi_0 + \mu^0_{104}\psi_1^{\dagger}\psi_1\psi_4\psi_0^{\dagger}\psi_0 + \\&\nu^0_{410}\psi_1^{\dagger}\psi_1\psi_4^{\dagger}\psi_4\psi_0^{\dagger} + \nu^0_{041}\psi_1^{\dagger}\psi_4^{\dagger}\psi_4\psi_0^{\dagger}\psi_0 +\\& \nu^0_{104}\psi_1^{\dagger}\psi_1\psi_4^{\dagger}\psi_0^{\dagger}\psi_0.
	 \end{split}
	 \end{equation}
Notice the presence of two kinds of coefficients: The coefficients related to number preserving operators ($\alpha, \beta, \xi, \mu$), from now on {\em np-coefficients}, and the coefficients related to non number preserving operators ($\gamma, \eta, \chi, \nu, \theta, \bar{\theta}$), from now on {\em nnp-coefficient}s. We can set a correspondence between pairs of coefficients in the two sets, by associating coefficients of terms that are one (proportional to) the adjoint of the other. 
The pairs of associated coefficients are then $(\alpha,\gamma)$, $(\beta,\eta)$, $(\xi,\chi)$, and $(\mu,\nu)$. The coefficients $\theta$ and $\bar{\theta}$ do not have a corresponding np-coefficient.

We exactly proceed as in Sec.~\ref{sec:ab} and we obtain a set of independent unitary equations. In this case the non null coefficients are
	\[ \alpha_1, \alpha_0, \alpha_4, \gamma_1, \gamma_0, \gamma_4, \beta_{10}, \beta_{40}, \eta_{10}, \eta_{40}, \mu_{410} \ \text{and} \ \nu_{410}. \]
From now on we will refer to $\mu_{410}$ as $\mu$ and to $\nu_{410}$ as $\nu$, since they are the only non null coefficients of terms of degree five. We notice that, differently from the system in Eq.~\eqref{sis:a-b-ab}, here only coefficients related to specific combinations of indices are non null.
	
Moreover, we notice that we can divide the whole set of equations into three mutually disjoint subsets. The first subset contains equations that are valid under the exchange of associated coefficient pairs. The second subset consists of equations that are invariant under the exchange of associated coefficient pairs, and the third subset collects the remaining equations.
	
We list now the three subsets of equations. As just said, this first subset is valid also substituting respectively $\alpha$, $\beta$ and $\mu$ coefficients with $\gamma$, $\eta$ and $\nu$ coefficients.
	\begin{align*}
	&\alpha_x\beta_{y0}=0 && \alpha_x\eta_{y0}=0 &\\
	& \alpha_x\mu=0 && \alpha_x\nu=0&\\
	&\alpha_0\beta_{x0}-\beta_{y0}\alpha_0=0 && \beta_{10}^2-\alpha_0\mu=0 &\\
	& \beta_{x0}\mu-\beta_{y0}\mu=0 && \beta_{x0}\nu+\eta_{y0}\mu=0&\\
	&\beta_{10}\eta_{10}+\alpha_0\nu+\beta_{10}\nu=0 && \beta_{x0}^*\eta_{x0}+\nu\alpha_0^*=0 &\\
	& \beta_{x0}^*\eta_{x0}+\mu^*\gamma_0=0 && \beta_{x0}\nu^*-\mu\eta_{y0}=0&\\
	& |\beta_{x0}|^2+\alpha_0\mu^*+\beta_{x0}\mu^*=0 && &\\
	&\beta_{x0}\beta_{y0}^*+\beta_{x0}\alpha_0^*+\alpha_0\beta_{y0}^*=0&& &\\
	&\beta_{x0}\mu^*+\beta_{y0}^*\mu+|\mu|^2=0&& &
	\end{align*}
where $x,y$ are neighborhoods indices. 
	
The second subset is 
	\begin{align*}
	&\sum_i (|\alpha_i|^2 + |\gamma_i|^2)=1 && \sum_i \alpha_i\gamma_i=0 &\\
	& \alpha_1\alpha_4+\alpha_4\gamma_1=0 &&\mu\nu=0  &\\
	& \alpha_4\eta_{x0}+\gamma_4\beta_{x0}=0 &&\alpha_1\alpha_4^*+\gamma_1\gamma_4^*=0  &\\
& \alpha_1\eta_{40}+\gamma_1\beta_{40}=0 &\\
	&\alpha_0\nu+\gamma_0\mu+\beta_{40}\eta_{10}+\beta_{10}\eta_{40}=0\\
	& \alpha_0\alpha_4^*+\alpha_1\alpha_0^*+\gamma_0\gamma_4^*+\gamma_1\gamma_0^*=0\\
	&\alpha_0\gamma_4+\alpha_1\gamma_0+\gamma_0\alpha_4+\gamma_1\alpha_0 = 0\\
	& \beta_{x0}\eta_{x0}+\alpha_0\eta_{x0}+\gamma_0\beta_{x0}=0\\
	&\alpha_0\beta_{x0}^*+\gamma_0\eta_{x0}^*+\eta_{x0}\gamma_0^*\\
	&+\eta_{x0}\eta_{x0}^*+\beta_{x0}\alpha_0^*+\beta_{x0}\beta_{x0}^* = 0 \\
	&\alpha_0\mu_{410}^*+\gamma_0\nu_{410}^*+\beta_{40}\beta_{10}^*\\
	&+\beta_{10}\beta_{40}^*+\beta_{10}\mu_{410}^*+\eta_{10}\eta_{40}^*\\
	&+\eta_{10}\nu_{410}^*+\eta_{40}\eta_{10}^*+\mu_{410}\alpha_0^*\\
	&+\mu_{410}\beta_{40}^*+\nu_{410}\gamma_0^*+\nu_{410}\eta_{40}^* = 0&
	\end{align*}
The third subset is 
	\begin{align*}
	&\alpha_0\eta_{40}+\beta_{10}\gamma_0=0 &\beta_{40}\eta_{40}-\mu_{410}\gamma_0=0  &\\
	&\eta_{40}\eta_{40}-\nu_{410}\gamma_0 = 0&\eta_{10}\alpha_0^*+\gamma_0\beta_{40}^* = 0 &\\
	&\eta_{10}\alpha_0+\eta_{10}\beta_{40}+\gamma_0\beta_{40} = 0 &\eta_{40}\alpha_0^*-\gamma_0\beta_{10}^* = 0  &\\
	& \beta_{40}\beta_{40}^*+\mu_{410}\alpha_0^*+\mu_{410}\beta_{40}^* = 0  &
	\end{align*}
	
Again, we can identify three families of solutions of the above system, grouped by the subset of non-null coefficients of terms of degree five. Indeed, since $\mu\nu=0$ we can have $\mu=0$, $\nu=0$ or $\mu=\nu=0$. Exploiting the polar notation for complex coefficients as in Sec.~\ref{sec:ab}, for the nnp-coefficients we have
\begin{align*}
&\gamma_x = |\gamma_x| e^{i \omega_x}, &&\eta_{xy} = |\eta_{xy}| e^{i \phi_{xy}}.
\end{align*}
Here we report the families of solutions.
\begin{enumerate}
\item $\mu=\nu= 0$: Only the coefficients $\alpha_0, \alpha_1, \alpha_4, \gamma_0, \gamma_1, \gamma_4 $ can be non-null. The conditions reduce to two sets of unitarity conditions for Quantum Walks.
\item ${\mu=0; \nu\neq 0}$: The non-null coefficients are $\gamma_0, \eta_{10}, \eta_{40}$. The unitarity conditions are
\begin{subequations}
	\begin{equation}
	\gamma_0\gamma_0^* = 1,
	\end{equation}
	\begin{equation}\label{eq:condEta2}
	2\cos(\phi_{{ij}} - \omega_{0}) = - |\eta_{ji}|,
	\end{equation}
	\begin{equation}\label{eq:condEta}
	\eta_{10}=\eta_{40}=:\eta = |\eta|e^{\phi},
	\end{equation}
	where without loss of generality we set $\gamma_0=1$. We consequently have
	\begin{equation}\label{eq:condtEta1}
	2\cos(\phi) = - |\eta|.
	\end{equation}
\end{subequations}
\item ${\nu=0; \mu\neq 0}$: The non-null coefficients are $\alpha_0, \beta_{10}, \beta_{40}$. The unitarity conditions are
\begin{subequations}
	\begin{equation}
	\alpha_0\alpha_0^* = 1,
	\end{equation}
	\begin{equation}\label{eq:condBeta2}
	2\cos(\varphi_{{ij}} - \theta_{0}) = - |\beta_{ji}|,
	\end{equation}
	\begin{equation}\label{eq:condBeta}
	\beta_{10}=\beta_{40}=: \beta =|\beta|e^{\varphi},
	\end{equation}
	where we set without loss of generality $\alpha_0=1$. We consequently have
	\begin{equation}\label{eq:condBeta2.1}
	2\cos(\varphi) = - |\beta|.
	\end{equation}
\end{subequations}
\end{enumerate}

In Tab.~\ref{tab:2} we report a synoptic summary of the above families of solutions.
	\begin{table}[h]
		\centering
		\begin{tabular*}{\columnwidth}{@{\extracolsep{\fill} }lll}
			\hline 
			${\mu=\nu= 0}$ &${\mu=0; \nu\neq 0}$ & ${\nu=0; \mu\neq 0}$ \\
			
			\hline 
			& $\gamma_0, \eta_{10}, \eta_{40} \neq 0$  & $\alpha_0, \beta_{10}, \beta_{40} \neq0$  \\ 
			
			& $\eta_{10}=\eta_{40}$ & $\beta_{10}=\beta_{40}$ \\ 
			
			Linear case & $\gamma_0\gamma_0^* = 1$& $\alpha_0 \alpha_0^*=1$ \\ 
			& $\nu\gamma_0 = \eta^2$  & $\mu\alpha_0 = \beta^2$  \\
			\rule[-8 pt]{0 pt}{0 pt}
			& $2\cos(\phi - \omega_{0})=- |\eta|  $ & $2\cos(\varphi - \theta_{0})=- |\beta|  $\\
			\hline 
		\end{tabular*}
		\caption{Families of solutions for the automata on Cayley graph related to $\left\langle a | a^5 \right\rangle$.}
		\label{tab:2}
	\end{table}

Differently from the case of $\mathbb Z_2\times\mathbb Z_2$, a non-linear automaton with a trivial non-linear sector is allowed if it includes both number preserving and non number preserving terms, as in the first family of solutions.

We notice that both the non-trivial families of non-linear solutions resemble very closely the non-trivial families of solutions in Tab.~\ref{tab:1}. For instance, we recover condition \ref{eq:condBeta1} if we choose $\theta_{0}=\omega_{0}=0$. Since the trilinear $\eta$ and $\beta$ coefficients are equal, the constraint on relative phases we found in Sec.~\ref{sec:ab}, see Tab.~\ref{tab:1}, is useless and does not appear. 

Notably, we point out that the second and the third families are connected to each other in a very natural way: Every number preserving solution corresponds to a non number preserving solution obtained by a global flipping transformation, i.e.~$\prod_{g\in \mathbb Z_5}(\psi_g+\psi_g^\dag)$, which maps $\psi_g\mapsto\psi_g^\dag$ (and vice versa). We conclude that non number preserving automata on the Cayley graph corresponding to $\left\langle a | a^5 \right\rangle$ are trivially obtained from the family of number preserving ones.
\section{Evolution operator}\label{sec:matrixev}
In this section we analyze the evolution provided by two non-linear unitary automata, one for each of the cases that we analyzed. To this end, we express the unitary evolution in a suitable basis, in order to have a visual interpretation of the eigenstate structure. Then, we compare the interacting evolutions with the corresponding (free) linear automata, described by the QW corresponding to the linear terms alone, as discussed in section \ref{sec:UnivQW}.
\subsection{Ordered basis and block structure}\label{subsec:basis}
Using the Jordan-Wigner Transform (JWT) representation, we explicitly calculate the evolution operator $U$ such that $\psi_x^{\prime}=U(\psi_x)U^\dag$. Indeed, using the JWT \cite{Jordan1928}, states and transformations of a system of $N$ fermionic modes are respectively mapped into vectors and operators of $N$ qubits. For this purpose, we choose a proper ordered basis $\{\left|e_i\right\rangle\}_{i=1}^{2^N}$ of the Fermionic algebra, where $N$ is the number of lattice sites, and we calculate the matrix elements $U_{ij}=\left\langle e_i|U|e_j \right\rangle$.

Creation and annihilation operators are then represented as
\begin{align}
&J(\psi_j)=\mathbb I_1\otimes\ldots\otimes\mathbb I_{j-1}\otimes\sigma^-_j\otimes\sigma^z_{j+1}\otimes\ldots\otimes\sigma^z_{N},\\
&J(\psi_j^\dag)=\mathbb I_1\otimes\ldots\otimes\mathbb I_{j-1}\otimes\sigma^+_j\otimes\sigma^z_{j+1}\otimes\ldots\otimes\sigma^z_{N},
\end{align}
where $\psi_j \xmapsto{JWT} J(\psi_j)$. With these elements we can recover the whole Fermionic algebra in terms of qubit states.

Finally, we identify a basis in the Hilbert space of the $N$ qubits by defining $|\Omega\rangle:=|0\ldots0\rangle$ the unique common eigenvector of the operators $\psi^\dag_i\psi_i$ with all the eigenvalues $0$, and then positing 
\begin{align}
|s_1\ldots s_N\rangle:=(\psi^\dag_N)^{s_N}\ldots(\psi^\dag_1)^{s_1}|\Omega\rangle,\quad s_j\in\{0,1\}.
\label{eq:fermibasis}
\end{align}
Notice that Fermionic theory prescribes the parity superselection rule \cite{DarianoetAl14,DarianoManessi14}. Pure states then correspond to rank one density matrices whose support is a superposition of vectors as in Eq.~\eqref{eq:fermibasis}.  Since even and odd elements cannot be combined, the superpositions of vectors can only involve elements of the basis $\{|s_1\ldots s_N\rangle\}_{s_j\in\{0,1\}}$ with the same parity $s_1\oplus\ldots\oplus s_N$, where $\oplus$ denotes sum modulo 2.

\subsection{Case $\mathbb Z_2\times\mathbb Z_2$}\label{subsec:Uab}
In this section we explicitly calculate the evolution operator $U$ for the solutions in Tab.~\ref{tab:1}, on the Cayley graph corresponding to $\left\langle a, b | a^2, b^2, (ab)^2 \right\rangle$. We chose the ordered basis as described in Subsec.~\ref{subsec:basis}. In the case of the Cayley graph $\left\langle a, b | a^2, b^2, (ab)^2 \right\rangle$, sticking to the ordering $(\psi_a, \psi_b, \psi_e, \psi_{c})$, we obtain the following ordered basis in the Hilbert space:
\begin{enumerate}
\item Even sector
\begin{enumerate}
\item Vacuum: $\left|0000\right\rangle=|\Omega\rangle$;
\item Two-particles:
\begin{align*}
&\left|0011\right\rangle, \left|0101\right\rangle, \left|0110\right\rangle,\\
& \left|1100\right\rangle, \left|1010\right\rangle, \left|1001\right\rangle;
\end{align*}
\item Totally excited state: $\left|1111\right\rangle=:|\bar{\Omega} \rangle$.
\end{enumerate}
\item Odd sector
\begin{enumerate}
\item Single particle: 
\begin{align*}
\left|1000\right\rangle, \left|0100\right\rangle, \left|0010\right\rangle, \left|0001\right\rangle;
\end{align*}
\item Three particles:
\begin{align*}
\left|1110\right\rangle,\left|1101\right\rangle,\left|1011\right\rangle,\left|0111\right\rangle.
\end{align*}
\end{enumerate}
\end{enumerate}
The vacuum and the totally excited states are invariant under the action of a number preserving evolution operator.

The two nontrivial families of solutions, see Tab.~\ref{tab:1} in Sec.~\ref{sec:ab} determine two different families of operators $U$. 
	\subsubsection{Odd sector: Single particle/hole sector}
	The expression to be evaluated is: 
	\begin{equation}\label{eq.singlepart}
	\langle 1_i | U | 1_j  \rangle = \langle \Omega | \psi_i U \psi_j^{\dagger} | \Omega  \rangle = \langle \Omega | \psi_i  (\psi_j^{\dagger})^{\prime} | \Omega  \rangle.
	\end{equation}
	In order to avoid ambiguities, we indicate with $\psi_{x_j}$ the field operator obtained by the action of the group generator $x$ on the starting field operator $\psi_j$ in the following. We report here the two different evolution operators for the two non-trivial families of solutions.
\begin{enumerate}
	\item Solutions with only one coefficient $\alpha_{x}\neq 0$.
	
	The matrix elements are 
	\begin{align*}
	\langle 1_i | &U | 1_j  \rangle =  \langle \Omega | \psi_i (\alpha_x\psi_{x_j}+\beta\psi_{t_j}^{\dagger}\psi_{t_j}\psi_{x_j}\\ &
	+\beta\psi_{s_j}^{\dagger}\psi_{s_j}\psi_{x_j}+ \gamma_{tsx}\psi_{t_j}^{\dagger}\psi_{t_j}\psi_{s_j}^{\dagger}\psi_{s_j}\psi_{x_j})^{\dagger}| \Omega  \rangle
	\end{align*}
	Since every non-linear term includes number operators, their contribution vanishes, as they annihilate the vacuum vector $|\Omega\rangle$. Thus,
	\begin{align}\label{eq:fam1}
	&\langle 1_i | U | 1_j  \rangle =  \langle \Omega | \psi_i \alpha_x^*\psi_{x_j}^{\dagger}| \Omega \rangle  =\alpha^*_x\delta_{i,x_j}.
	\end{align}
	\item Solutions with two coefficients $\alpha_{x},\alpha_y\neq0$.
	
	In this case the matrix elements are 
	\begin{align*}
	&\langle 1_i | U | 1_j  \rangle =  \langle \Omega | \psi_i (\alpha_x\psi_{x_j}+\alpha_y\psi_{y_j} \\&+\beta_{xy}\psi_{x_j}^{\dagger}\psi_{x_j}\psi_{y_j}+\beta_{yx}\psi_{y_j}^{\dagger}\psi_{y_j}\psi_{x_j})^{\dagger}| \Omega  \rangle.
	\end{align*}
	Again, terms containing number operators give null contributions, then we have
	\begin{align}\label{eq:fam2}
	&\langle 1_i | U | 1_j  \rangle =  \langle \Omega | \psi_i (\alpha_x^*\psi_{x_j}^{\dagger}+\alpha_y^*\psi_{y_j}^{\dagger})| \Omega  \rangle 
	\end{align}
\end{enumerate}
The above Eq.~\eqref{eq:fam1} and Eq.~\eqref{eq:fam2} have the same values. Indeed, by definition we have that $x\neq y$ for every pair $x,y$; consequently either $x_j=i$ or $y_j= i$ and for fixed $i$ and $j$ only one of the two terms $\alpha_x^*\psi_{x_j}^{\dagger}, \alpha_y^*\psi_{y_j}^{\dagger}$ gives a non null contribution to the element in Eq.~\eqref{eq:fam2}. As a result we have the same sub-matrices for both the families.

The other states of the odd sector are the three particle states. We denote $| 3_{lmn}  \rangle  :=\psi_l^\dag \psi_m^\dag \psi_n^\dag| \Omega \rangle $ and we can exploit the relation
\begin{align*}\label{eq:oddab}
\langle 3_{ijk} | U | 3_{lmn}  \rangle &= \langle \Omega |\psi_i\psi_j\psi_k U \psi_l^\dag \psi_m^\dag \psi_n^\dag| \Omega \rangle \\
&=(-1)^{\varphi(p)+\varphi(q)} \langle \bar{\Omega} | \psi_p^\dag U \psi_q| \bar{\Omega}  \rangle,
\end{align*}
where 
\begin{align*}
|\bar\Omega\rangle=(-1)^{\varphi(p)}\psi^\dag_p\psi^\dag_l\psi^\dag_m\psi^\dag_n|\Omega\rangle=(-1)^{\varphi(q)}\psi^\dag_q\psi^\dag_k\psi^\dag_j\psi^\dag_i|\Omega\rangle.
\end{align*}
Consequently, the block representing $U$ in the three-particle sector can be evaluated as a single-hole sector.

\subsubsection{Even sector: Two particles sector}
It is clear that the only non-trivial transitions in the even sector are among two particles states because vacuum and totally excited states are invariant under the evolution. In this case the elements to be evaluated are
	\begin{equation}\label{key}
	\begin{split}
	\langle 2_{i,j} | U | 2_{k,l}  \rangle &= \langle \Omega | \psi_i \psi_j U \psi_k^{\dagger}\psi_l^{\dagger} | \Omega  \rangle\\
	& = \langle \Omega | \psi_i \psi_j   (\psi_k^{\dagger})^{\prime}(\psi_l^{\dagger})^{\prime} | \Omega  \rangle.
	\end{split}
	\end{equation}
	Choosing a proper normal ordering we can find a block structure for these sub-matrices as shown in Appendix~\ref{app:B}. We recall the normal ordering for the even sector:
	$\left|0011\right\rangle, \left|0101\right\rangle, \left|0110\right\rangle, \left|1100\right\rangle, \left|1010\right\rangle, \left|1001\right\rangle$. Again, the matrix elements for the two families are:

\begin{enumerate}
 \item Solutions with  with only one coefficient $\alpha_{x}\neq 0$.
	
The matrix elements in this case are 
\begin{equation*}
\begin{split}
\langle 2_{ij} | U | 2_{kl}  \rangle =\langle \Omega | \psi_i\psi_j(\alpha_x\psi_{x_k}+\beta\psi_{t_k}^{\dagger}\psi_{t_k}\psi_{x_k}\\ +\beta\psi_{s_k}^{\dagger}\psi_{s_k}\psi_{x_k} + \gamma_{tsx}\psi_{t_k}^{\dagger}\psi_{t_k}\psi_{s_k}^{\dagger}\psi_{s_k}\psi_{x_k})^{\dagger}\\
(\alpha_x\psi_{x_l}+\beta\psi_{t_l}^{\dagger}\psi_{t_l}\psi_{x_l}+\beta\psi_{s_l}^{\dagger}\psi_{s_l}\psi_{x_l}\\ + \gamma_{tsx}\psi_{t_l}^{\dagger}\psi_{t_l}\psi_{s_l}^{\dagger}\psi_{s_l}\psi_{x_l})^{\dagger}| \Omega  \rangle\\
\end{split}
\end{equation*}
\item Solutions with two coefficients $\alpha_{x},\alpha_y\neq0$.
	
In this case the matrix elements are 
\begin{equation*}
\begin{split}
\left\langle 2_{ij} | U | 2_{kl} \right \rangle = &\langle \Omega | \psi_i \psi_j(\alpha_x\psi_{x_k}+\alpha_y\psi_{y_k} +\beta_{xy}\psi_{x_k}^{\dagger}\psi_{x_k}\psi_{y_k} \\& +\beta_{xy}\psi_{x_k}^{\dagger}\psi_{x_k}\psi_{y_k})^{\dagger} (\alpha_x\psi_{x_l}+\alpha_y\psi_{y_l} \\&+\beta_{xy}\psi_{x_l}^{\dagger}\psi_{x_l}\psi_{y_l} +\beta_{xy}\psi_{x_l}^{\dagger}\psi_{x_l}\psi_{y_l})^{\dagger}| \Omega  \rangle.
\end{split}
\end{equation*}
\end{enumerate}
The block-structure of $U$ is the following
\begin{align}
U=
\begin{pmatrix}
1&0&0&0&\vline&0&0\\
0&A&B&0&\vline&0&0\\
0&B&A&0&\vline&0&0\\
0&0&0&1&\vline&0&0\\
\hline
0&0&0&0&\vline&S&0\\
0&0&0&0&\vline&0&T
\end{pmatrix},
\end{align}
where $A,B \in \mathcal{M}(3, \mathbb C)$, $S,T \in \mathcal{M}(4, \mathbb C)$, with $\mathcal{M}(n, \mathbb C)$ denoting the set of square complex matrices $n\times n$, and the remaining zeros denoting suitable rectangular matrices with null elements.
We explicitly report the blocks $A,B,S,T$ of $U$ for the two families of solutions of table \ref{tab:1} in Appendix~\ref{app:B}. 

\subsection{Case $\mathbb Z_5$}\label{subsec:Ua}
Here we explicitly calculate operator $U$ for the case of Cayley graphs corresponding to $\left\langle a | a^5 \right\rangle$. The two families of solutions are shown in Tab.~\ref{tab:2} in Sec~\ref{sec:a}. 
We notice that in the two cases the evolved field is a linear combination of terms that are exclusively non-number preserving or number preserving, respectively. 

Vacuum and totally excited states are not invariant states for the non-number preserving evolution. Despite these differences, we can connect the two evolutions. Indeed, non null coefficients for the two families are related: For every non-null coefficient of the number preserving (np) family of solutions, the corresponding coefficient in the non-number preserving (nnp) case is non-null, and vice-versa. 

Let $U^\mathrm{(np)}$ denote the evolution operator of the general np solution, and $U^\mathrm{(nnp)}$ that of the general nnp solution. We know that for every np automaton, a nnp one can be obtined by a global flipping transformation. So 
	\[ \forall\, U^{(\mathrm{nnp})}, \exists\, U^\mathrm{(np)}: U^{(\mathrm{nnp})} = F U^{(\mathrm{np})} \]
where  $F$ is the global flipping operator. Consequently, 
	\[ U^\mathrm{(nnp)}\left| \Omega \right \rangle = F U^\mathrm{(np)}\left| \Omega \right \rangle = \left| \bar{\Omega}\right \rangle. \]
	
Here we present the matrix form M of the evolution operators for the number preserving family, see Tab.~\ref{tab:2}. The matrix form N for the other family can then be obtained by $N_{ij} = M_{32-i, j}$. We choose the normal ordering $\{\psi_1, \psi_0, \psi_4, \psi_3, \psi_2\}$ for the basis of the Fermionic algebra and proceed using the JWT as in the previous Subsec.~\ref{subsec:Uab}. The ordered basis of the qubit space is
\begin{enumerate}
\item Even sector
\begin{enumerate}
\item Vacuum: $\left|00000\right\rangle=|\Omega\rangle$;
\item Two-particles:
\begin{align*}
&\left|11000\right\rangle,\ \left|10100\right\rangle,\ \left|10010\right\rangle,\ \left|10001\right\rangle,\ \left|01100\right\rangle,\\ 
&\left|01010\right\rangle,\ \left|01001\right\rangle,\ \left|00110\right\rangle,\ \left|00101\right\rangle,\ \left|00011\right\rangle;
\end{align*}
\item Four particles: 
\begin{align*}
&\left|11110\right\rangle,\ \left|11101\right\rangle,\ \left|11011\right\rangle,\ \left|10111\right\rangle,\ \left|01111\right\rangle.
\end{align*}
\end{enumerate}
\item Odd sector
\begin{enumerate}
\item Single particle: 
\begin{align*}
&\left|10000\right\rangle,\ \left|01000\right\rangle,\ \left|00100\right\rangle,\ \left|00010\right\rangle,\ \left|00001\right\rangle;
\end{align*}
\item Three particles:
\begin{align*}
&\left|11100\right\rangle,\ \left|11001\right\rangle,\ \left|10011\right\rangle,\ \left|00111\right\rangle,\ \left|01110\right\rangle, \\
&\left|01011\right\rangle,\ \left|10101\right\rangle,\ \left|11010\right\rangle,\ \left|01101\right\rangle,\ \left|10110\right\rangle;
\end{align*}
\item Totally excited state: $\left|11111\right\rangle.$
\end{enumerate}
\end{enumerate}

We enumerate the chosen basis by an index $i\in I, \ \text{with} \ I\subset\mathbb{N}$, and calculate the submatrix $A$ of transition amplitudes between single particle states:
\[ 	(A)_{ij} = \alpha_0^*\delta_{i,j}, \]
and the submatrix $B$ of transition amplitudes between three particles states is
\begin{itemize}
	\item $(B)_{ij} = (\alpha_0+\beta)(\alpha_0+2\beta+\mu)\delta_{ij}$ for $i=j \in[22, 26]$ 
	\item $(B)_{ij} = (\alpha_0+2\beta+\mu)^2\delta_{ij}$ for $i=j \in[27, 31].$ 
\end{itemize}
Then we have the even sector. The submatrix $C$ of transition amplitudes between single hole states is
\[ (C)_{ij} = (\alpha_0+2\beta +\mu)\delta_{i,j}, \]
and the submatrix $D$ of transition amplitudes between two particle states is
\[ (D)_{ij} = (\alpha_0^*)^2\delta_{i,j}. \]

\subsection{Discriminating linear and non-linear evolutions}\label{subsec:discri}
In this section we briefly analyse the phenomenological aspects of the nonlinear unitary evolutions. As it is clear in Appendix~\ref{app:B}, from the point of view of patterns of localized excitations, our non-linear evolutions are not qualitatively different from QW-like typical transitions.

We cannot locally discriminate between a non-linear evolution and the corresponding linear one, as long as we prepare localized states. The discrimination problem between linear and non-linear evolutions thus requires delocalized states, with the relevant dynamical information encoded in the different phase shifts affecting vectors representing localized states. One can check this statement by carefully looking at the optimal states for discrimination of unitary evolutions.

If we observe the evolution of a basis of localized excitations, we see that the same transitions occur between localized configurations both in the linear case and in the nonlinear case, however with different phases of the transition amplitudes. The comparison between the evolution of a non-linear automaton and the corresponding linearized one can be cast in terms of a discrimination problem between two Fermionic unitary operators. 

Following the analysis of Ref.~\cite{Paris01}, let $U_0, U_1$ respectively be a linear and a nonlinear evolution operator we want to discriminate between, it is not restrictive to discriminate between the identity operator $I=\tilde{U^{}}_0 $ and $\tilde{U^{}}_1= U_0^{\dagger}U_1$ instead. Since the operator $\tilde{U^{}}_1= U_0^{\dagger}U_1$ is a combination of Fermionic evolution operators, it can only be either parity preserving or parity flipping. We analyse the two cases separately.

When $\tilde{U^{}}_1= U_0^{\dagger}U_1$ is parity flipping, $\tilde{U}_0$ and $\tilde{U}_1$ are perfectly discriminable. Indeed, let $| \Psi_E \rangle$ an even state and let $P_E$ the projector on the even subspace. We have
\[ \langle \Psi_E | \tilde{U}_i^\dag P_E \tilde{U}_i | \Psi_E \rangle= \delta_{i,0}. \]
So we have a perfectly discriminating procedure between the two operators.

Otherwise $\tilde{U}_1= U_0^{\dagger}U_1$ is parity preserving. Again following Ref.~\cite{Paris01}, the most general procedure to discriminate a unitary (in our case parity preserving) map $\mathcal{U}$ from the identity map is applying the unknown transformation on one side of an entangled bipartite state $\Phi$, which gives
\begin{align*}
|\Psi_i\rangle := \tilde U_i \otimes I |\Phi\rangle,
\end{align*}
The discrimination probability is then a decreasing function of the overlap between the vectors $ |\Psi_i\rangle $, which is given by
\begin{align*}
|\langle\Psi_0|\Psi_1\rangle|=|\langle \Phi | U \otimes I |\Phi\rangle| = |\mathrm{Tr}[U\Phi \Phi^\dag]| = |\mathrm{Tr}[U\rho]|,
\end{align*}
where $\rho=\operatorname{Tr}_2[|\Phi\rangle\langle\Phi|]$. Now, diagonalising $U$, we obtain
\begin{align}
\mathrm{Tr}[U\rho] = \sum_j e^{i \theta_j} \langle \eta_j | \rho | \eta_j \rangle,
\label{eq:convcomb}
\end{align}
with $|\eta_j\rangle$ eigenvector of $U$ corresponding to the eigenvalue $e^{i \theta_j}$.

The eigenvalues of $U$ are distributed over the unit circle in the complex plane $\mathbb C$, and can be geometrically represented as the vertices of a polygon. The overlap $|\langle\Psi_0|\Psi_1\rangle|$ amounts to the distance of the centre of the circumference from the inscribed polygon: If the centre is inside the polygon we can perfectly discriminate the two unitaries.

Optimal strategies correspond to purifications of any state $\rho$ with optimal weights $\langle \eta_j | \rho | \eta_j \rangle=p_i$, such that the expression in Eq.~\eqref{eq:convcomb} amounts to the minimum distance point of the polygon from the origin. In the usual quantum case, we can in fact always achieve the optimal discrimination with a {\em local} procedure, corresponding to the choice $\rho=|\chi\rangle\langle\chi|$, and $|\chi\rangle := \sum \sqrt{p_j} |\eta_j\rangle$.

In the Fermionic case the situation is slightly more complicate. Indeed, every state $\rho$  corresponding to the minimum distance point of the polygon from the origin gives the optimal strategy also in this case, however with a caveat. Indeed, if the minimum distance point is a convex combination involving eigenvalues with both even and odd eigenvectors, then the preparation for the optimal state cannot be local, and one needs an ancillary system to purify a mixture of even and odd states. 

The optimal {\em local} discrimination procedure, on the other hand, is obtained by the same geometric construction as above, however restricting attention to the two polygons whose vertices are those of eigenvalues corresponding to {\em even} and {\em odd} eigenvectors, respectively. Thus, the optimal procedure does not require entanglement if there exists a set $J$ of eigenvalues related to eigenvectors of given parity, whose distance from the origin is the same as the full polygon. 

In this case, again we can find an optimal local state $|\chi\rangle$ given by the even (or odd) superposition $|\chi\rangle = \sum \sqrt{p_j} |\eta_j\rangle$, with $|\eta_j\rangle$ eigenvectors related to eigenvalues in $J$. 


Then, we need to calculate $U_0^{\dagger}U_1$. In App.~\ref{app:B} we give the explicit matrix form of the evolution operator $U_1$ for the third family of solutions of the case $\left\langle a, b | a^2, b^2, (ab)^2 \right\rangle$ and we can easily calculate $U_0$, which trivially consists in $U_1$ stripped of all its non-linear coefficients. In App.~\ref{app:C} we explicitly calculate eigenvalues of $\tilde{U}_1$ for the two nontrivial families of solutions for the Cayley graph of $\left\langle a, b | a^2, b^2, (ab)^2 \right\rangle$. We find that the optimal discrimination strategy can always avoid entanglement, and the optimal discrimination probability is given by
\begin{align*}
&p_\mathrm{succ}=\sin\frac \Theta2,\ \Theta:=\arg\lambda,\\
&\lambda:=((\alpha_x^*)^2+ (\alpha_y^*)^2)((\alpha_x^*)^2+ (\alpha_y^*)^2 + \beta_{yx}^*\alpha_x^* + \beta_{xy}^*\alpha_y^*).
\end{align*}
Then, the perfect discriminability condition is
\begin{align*}
\Theta=\pi.
\end{align*}

\section{Summary}\label{sec:conclu}
We analysed two simple but paradigmatic non-linear FCAs in order to find conditions for their unitary evolution. In order to make FCAs relevant tools for the recontruction of the non-trivial interacting QFT dynamics, we preliminary imposed homogeneity and locality of interaction. Thank to these requirements we were able to describe the neighbourhood structure of FCAs with Cayley graphs. We chose the groups $\left\langle a, b | a^2, b^2, (ab)^2 \right\rangle$
and $\left\langle a | a^5 \right\rangle$ and found the exact unitary solutions respectively for a number preserving evolution and for a non number preserving evolution. In each case, solutions can be arranged in three different families with notably analogies among them.

We found solutions that imply transitions among states that are not qualitatively different form a linear case. Nevertheless we highlighted that is always possible to discriminate these non-linear evolutions from the associated linear ones for a proper choice of the phases of the evolution coefficients.

These results represent the first attempt to find a universal procedure to identify unitarity conditions for a general non-linear FCA. Indeed, we already found two general prescriptions in this sense. Unitarity conditions for Quantum Walks must be valid for non-linear FCAs too. Indeed, in the present case studies we cannot have a non-linear FCA with a trivial linear sector. Moreover, we pointed out that not all the non-linear operators are allowed in a unitary evolution and we identify a kind of term that is not allowed.

\acknowledgments This publication was made possible through the support of a grant from the John Templeton Foundation under the project ID\# 60609 Causal Quantum Structures. The opinions expressed in this publication are those of the authors and do not necessarily reflect the views of the John Templeton Foundation.

\bibliography{bibliografia}

\begin{appendix}
\begin{widetext}
	\newpage
	\flushleft
\section{Matrix form of the evolution operators}\label{app:B}

\subsection{Solution $\psi'_l= \alpha_x \psi_{x_l} + \beta_{ix}\psi_{i_l}^\dag\psi_{i_l}\psi_{x_l}  + \beta_{jx}\psi_{j_l}^\dag\psi_{j_l}\psi_{x_l}  + \gamma_{ijx}\psi_{i_l}^\dag\psi_{i_l}\psi_{j_l}^\dag\psi_{j_l}\psi_{x_l} $}
\begin{itemize}
	\item Matrix S$ $\\
	with $x=e$: $ \begin{pmatrix}
	\alpha_e^* & 0 & 0& 0 \\
	0& \alpha_e^*&  0 & 0\\
	0 & 0 & \alpha_e^* & 0\\
	0 & 0 & 0 & \alpha_e^*
	\end{pmatrix} 
	;
	$
	with $x=a$: $\begin{pmatrix}
	0 & 0 & \alpha_a^*& 0 \\
	0& 0&  0 & \alpha_a^*\\
	\alpha_a^* & 0 & 0 & 0\\
	0 & \alpha_a^* & 0 & 0
	\end{pmatrix} 
	;$
	with $x=b$: $\begin{pmatrix}
	0 & 0 & 0& \alpha_b^* \\
	0& 0& \alpha_b^* &0\\
	0 & \alpha_b^* & 0 & 0\\
	\alpha_a^*& 0 & 0 & 0
	\end{pmatrix} 
	$.
	\\
	\vspace*{0.3in}
	\item Matrix T, with the same eigenvector as matrix $S$, with respect to the different cases $ $\\
	with $x=e$: $ \begin{pmatrix}
	z & 0 & 0& 0 \\
	0& z&  0 & 0\\
	0 & 0 & z & 0\\
	0 & 0 & 0 & z
	\end{pmatrix} 
	$
	with $x=a$: $\begin{pmatrix}
	0 & 0 & w& 0 \\
	0& 0&  0 & w\\
	w & 0 & 0 & 0\\
	0 & w & 0 & 0
	\end{pmatrix} 
	$
	with $x=b$: $\begin{pmatrix}
	0 & 0 & 0& t \\
	0& 0& t &0\\
	0 & t & 0 & 0\\
	t& 0 & 0 & 0
	\end{pmatrix} 
	$
	
	with eigenvalues $z, w, t$ and $z=\alpha_e + \beta_{ae} + \beta_{be} + \gamma_{abe}$, $w=\alpha_a + \beta_{ba} + \beta_{ea} + \gamma_{bea}$, $t=\alpha_b + \beta_{ab} + \beta_{eb} + \gamma_{eab}$.
	\\
	\vspace*{0.3in}
	\item Matrix A with $x=e, a, b$; eigenvalues and eigenvectors are trivial:
	$\begin{pmatrix}
	(\alpha_e^*)^2& 0& 0 \\
	0 & (\alpha_e^*)^2 + \beta^*\alpha_e^*& 0\\
	0& 0& (\alpha_e^*)^2 + \beta^*\alpha_e^* \\
	\end{pmatrix}; \quad
	\begin{pmatrix}
	0& 0 & 0 \\
	0 & (\alpha_a^*)^2 + \beta^*\alpha_a^* & 0 \\
	0 & 0& 0 \\
	\end{pmatrix}; \quad
	\begin{pmatrix}
	0& 0 & 0 \\
	0 & 0 & 0 \\
	0 & 0 & (\alpha_b^*)^2 + \beta^*\alpha_b^* \\
	\end{pmatrix} $.
	\\
	\vspace*{0.3in}
	\item Matrix B with $x=e, a, b$; eigenvalues and eigenvectors are trivial:
	$\begin{pmatrix}
	0& 0 & 0 \\
	0 & 0 & 0\\
	0& 0 & 0 \\
	\end{pmatrix}; \quad
	\begin{pmatrix}
	(\alpha_a^*)^2 + \beta^*\alpha_a^* & 0 & 0 \\
	0 & 0 & 0 \\
	0 & 0 & (\alpha_a^*)^2 \\
	\end{pmatrix}; \quad
	\begin{pmatrix}
	(\alpha_b^*)^2 + \beta^*\alpha_b^* & 0 & 0 \\
	0 & (\alpha_b^*)^2 & 0 \\
	0 & 0 & 0 \\
	\end{pmatrix}$.
\end{itemize}

We report now the matrix $V$ having eigenvectors (of the matrix form of the evolution operator) as columns. The case $x=e$ is trivial, being the evolution operator diagonal, while the eigenvectors are the same for the other two case .
	$\setcounter{MaxMatrixCols}{14}	
V=	\begin{pmatrix}
	0& 0& 0& 0&  0& 0& 1& 0& -1&  0&  0& 0&  0&  0\\
	0& 0& 0& 0& -1& 0& 0& 0&  0&  0&  0& 1&  0&  0\\
	0& 0& 0& 0&  0& 1& 0& 0&  0&  0&  0& 0&  0&  0\\
	0& 0& 0& 0&  0& 0& 1& 0&  1&  0&  0& 0&  0&  0\\
	0& 0& 0& 0&  1& 0& 0& 0&  0&  0&  0& 1&  0&  0\\
	0& 0& 0& 0&  0& 0& 0& 1&  0&  0&  0& 0&  0&  0\\
	0& 1& 0& 0&  0& 0& 0& 0&  0&  0& -1& 0&  0&  0\\
	1& 0& 0& 0&  0& 0& 0& 0&  0& -1&  0& 0&  0&  0\\
	1& 0& 0& 0&  0& 0& 0& 0&  0&  1&  0& 0&  0&  0\\
	0& 1& 0& 0&  0& 0& 0& 0&  0&  0&  1& 0&  0&  0\\
	0& 0& 0& 1&  0& 0& 0& 0&  0&  0&  0& 0&  0& -1\\
	0& 0& 1& 0&  0& 0& 0& 0&  0&  0&  0& 0& -1&  0\\
	0& 0& 1& 0&  0& 0& 0& 0&  0&  0&  0& 0&  1&  0\\
	0& 0& 0& 1&  0& 0& 0& 0&  0&  0&  0& 0&  0&  1
	\end{pmatrix}	
	$

\subsection{Solution $\psi'_l= \alpha_x \psi_{x_l} + \alpha_y \psi_{y_l} + \beta_{xy}\psi_{x_l}^\dag\psi_{x_l}\psi_{y_l}  + \beta_{yx}\psi_{y_l}^\dag\psi_{y_l}\psi_{x_l}$}
\begin{itemize}
	\item Matrix S with $x=e, a, b$:
	$\begin{pmatrix}
	\alpha_e^* & 0 & 0 & 0 \\
	0 & \alpha_e^* & 0 & 0 \\ 
	0 & 0 & \alpha_e^* & 0 \\
	0 & 0 & 0 & \alpha_e^*
	\end{pmatrix}, \quad
	\begin{pmatrix}
	0 & 0 & \alpha_a^*& 0 \\
	0& 0&  0 & \alpha_a^*\\
	\alpha_a^* & 0 & 0 & 0\\
	0 & \alpha_a^* & 0 & 0\\
	\end{pmatrix},\quad
	\begin{pmatrix}
	0 & 0 & 0 & \alpha_b^* \\
	0 & 0 & \alpha_b^* & 0 \\
	0 & \alpha_b^* & 0 & 0 \\
	\alpha_b^* & 0 & 0 & 0
	\end{pmatrix};$\\
	\item Matrix T\\
	with $(x,y)=(e,a)$ and $ (x,y)=(e,b) $:
	$\begin{pmatrix}
	\alpha_e +\beta_{ae} & 0 & 0 & 0 \\
	0 & \alpha_e +\beta_{ae} & 0 & 0 \\ 
	0 & 0 & \alpha_e +\beta_{ae} & 0 \\
	0 & 0 & 0 & \alpha_e +\beta_{ae}
	\end{pmatrix}, \quad
	\begin{pmatrix}
	\alpha_e +\beta_{be} & 0 & 0 & 0 \\
	0 & \alpha_e +\beta_{be} & 0 & 0 \\ 
	0 & 0 & \alpha_e +\beta_{be} & 0 \\
	0 & 0 & 0 & \alpha_e +\beta_{be}
	\end{pmatrix};$\\
	with $(x,y)=(a,e)$ and $ (x,y)=(a,b)$:
	$\begin{pmatrix}
	0 & 0 & \alpha_a +\beta_{ea}& 0 \\
	0& 0&  0 & \alpha_a +\beta_{ea}\\
	\alpha_a +\beta_{ea} & 0 & 0 & 0\\
	0 & \alpha_a +\beta_{ea} & 0 & 0
	\end{pmatrix},\quad
	\begin{pmatrix}
	0 & 0 & \alpha_a +\beta_{ba}& 0 \\
	0& 0&  0 & \alpha_a +\beta_{ba}\\
	\alpha_a +\beta_{ba} & 0 & 0 & 0\\
	0 & \alpha_a +\beta_{ba} & 0 & 0
	\end{pmatrix};$\\
	with $(x,y)=(b,e)$ and $ (x,y)=(b,a)$:
	$\begin{pmatrix}
	0 & 0 & 0 & \alpha_b+\beta_{be} \\
	0 & 0 & \alpha_b+\beta_{be} & 0 \\
	0 & \alpha_b+\beta_{be} & 0 & 0 \\
	\alpha_b+\beta_{be} & 0 & 0 & 0
	\end{pmatrix},\quad
	\begin{pmatrix}
	0 & 0 & 0 & \alpha_b+\beta_{ba} \\
	0 & 0 & \alpha_b+\beta_{ba} & 0 \\
	0 & \alpha_b+\beta_{ba} & 0 & 0 \\
	\alpha_b+\beta_{ba} & 0 & 0 & 0
	\end{pmatrix};$\\
	\item Matrix A with $(x,y)=(e,a); (e,b); (b,a)$:
	
	 $ \begin{pmatrix}
	(\alpha_e^*)^2& 0& \alpha_a^*\alpha_e^* \\
	0 & z& 0\\
	\alpha_e^*\alpha_a^*& 0& (\alpha_e^*)^2 \\
	\end{pmatrix}
$
	;
	$ \begin{pmatrix}
	(\alpha_e^*)^2 & \alpha_b^*\alpha_e^* & 0 \\
	\alpha_b^*\alpha_e^* & (\alpha_e^*)^2 & 0 \\
	0 & 0& w \\
	\end{pmatrix} 
	$; 
$ \begin{pmatrix}
	0& 0 & 0 \\
	0 & (\alpha_a^*)^2 & \alpha_a^*\alpha_b^* \\
	0 & \alpha_b^*\alpha_a^* & (\alpha_b^*)^2 \\
	\end{pmatrix} 
	$;
	\\
	with $z=(\alpha_e^*)^2 + (\alpha_a^*)^2 + \beta_{ea}^*\alpha_a^* +  \beta_{ae}^*\alpha_e^*$ and  $w=(\alpha_e^*)^2+(\alpha_b^*)^2+\beta_{eb}^*\alpha_b^*+\beta_{be}^*\alpha_e^*$\\
	\item Matrix B with $(x,y)=(e,a); (e,b); (b,a)$: \\ 
		
	$ \begin{pmatrix}
	(\alpha_a^*)^2& 0 & \alpha_a^*\alpha_e^* \\
	0 & 0 & 0\\
	\alpha_a^*\alpha_e^* & 0 & (\alpha_a^*)^2 \\
	\end{pmatrix} 
	$;
	$ \begin{pmatrix}
	(\alpha_b^*)^2 & \alpha_b^*\alpha_e^* & 0 \\
	\alpha_b^*\alpha_e^* & (\alpha_b^*)^2 & 0 \\
	0 & 0 & 0 \\
	\end{pmatrix} 
	$;
	$ \begin{pmatrix}
	z & 0 & 0 \\
	0 & (\alpha_b^*)^2 & \alpha_b^*\alpha_a^* \\
	0 & \alpha_b^*\alpha_a^* & (\alpha_a^*)^2 \\
	\end{pmatrix} 
	$.\\
	with $z=(\alpha_a^*)^2 + \beta_{ab}^*\alpha_b^* + (\alpha_b^*)^2 + \beta_{ba}^*\alpha_a^*$ 
\end{itemize}
We report the different $V, W$ matrices having eigenvectors as columns:
For $(x,y)=(e,a)$ and $(e,b)$ we have the same eigenvectors and the same $V$ matrix. For $(x,y)=(b,a)$ we have $W$ with $z=-\sqrt{((\beta_{ba} + \alpha_a)(\beta_{ba} + \alpha_b))}/(\beta_{ba} + \alpha_a)$:
	
	$\small \setcounter{MaxMatrixCols}{14}	
	V=	\begin{pmatrix}
	 0& 0& 0& 0& 0& 0& -1&  0& 0& 0& 0& 0& -1& 1\\	
	 0& 0& 0& 0& 0& 0&  0& -1& 0& 0& 0& 0&  1& 1\\
	 0& 0& 0& 0& 1& 0&  0&  0& 0& 0& 0& 0&  0& 0\\
	 0& 0& 0& 0& 0& 0&  1&  0& 0& 0& 0& 0& -1& 1\\
	 0& 0& 0& 0& 0& 0&  0&  1& 0& 0& 0& 0&  1& 1\\
	 0& 0& 0& 0& 0& 1&  0&  0& 0& 0& 0& 0&  0& 0\\
	 1& 0& 0& 0& 0& 0&  0&  0& 0& 0& 0& 0&  0& 0\\
	 0& 1& 0& 0& 0& 0&  0&  0& 0& 0& 0& 0&  0& 0\\
	 0& 0& 1& 0& 0& 0&  0&  0& 0& 0& 0& 0&  0& 0\\
	 0& 0& 0& 1& 0& 0&  0&  0& 0& 0& 0& 0&  0& 0\\
	 0& 0& 0& 0& 0& 0&  0&  0& 1& 0& 0& 0&  0& 0\\
	 0& 0& 0& 0& 0& 0&  0&  0& 0& 1& 0& 0&  0& 0\\
	 0& 0& 0& 0& 0& 0&  0&  0& 0& 0& 1& 0&  0& 0\\
	 0& 0& 0& 0& 0& 0&  0&  0& 0& 0& 0& 1&  0& 0\\
	\end{pmatrix}	
	$,
	$W=	\begin{pmatrix}
		 0& 0& 1&0&  0&  0&  0& 0&  0&  0& -1& 0&  0& 0\\
		 0& 0& 0& 0&  0& -1&  0& 0&  0&  0&  0& 0& -1& 1\\
		 0& 0& 0& 0& -1&  0&  0& 0&  0&  0&  0& 0&  1& 1\\
		 0& 0& 1& 0&  0&  0&  0& 0&  0&  0&  1& 0&  0& 0\\
		 0& 0& 0& 0&  0&  1&  0& 0&  0&  0&  0& 0& -1& 1\\
		 0& 0& 0& 0&  1&  0&  0& 0&  0&  0&  0& 0&  1& 1\\
		 0& 1& 0& 0&  0&  0&  0& 0&  0& -1&  0&  0&  0& 0\\
		 1& 0& 0& 0&  0&  0&  0& 0& -1&  0&  0& 0&  0& 0\\
		 1& 0& 0& 0&  0&  0&  0& 0&  1&  0&  0&  0&  0& 0\\
		 0& 1& 0& 0&  0&  0&  0& 0&  0&  1&  0& 0&  0& 0\\
		 0& 0& 0& -z&  0&  0&  0& 0&  0&  0&  0& z&  0& 0\\
		 0& 0& 0& 0&  0&  0& -1& 1&  0&  0&  0& 0&  0& 0\\
		 0& 0& 0& 0&  0&  0&  1& 1&  0&  0&  0& 0&  0& 0\\
		 0& 0& 0& 1&  0&  0&  0& 0&  0&  0&  0& 1&  0& 0\\
	\end{pmatrix}$.
\end{widetext}
\section{Explicit calculation of operator $ \tilde{U}_1 $} \label{app:C}
We calculate explicitly the matrix form of the operator $ \tilde{U}_1 $ for both the two nontrivial families of solutions for the graph associated with  $\left\langle a, b | a^2, b^2, (ab)^2 \right\rangle$. As made clear in Sec.~\ref{subsec:discri}, we are interested in the eigenvalues of this matrix in order to discriminate between a linear and a non-linear evolution. Matrix $U_1$ for these cases is presented in the previous section. 

We are looking for even eigenvalues in order to locally discriminate between a free and an interacting evolution so we now analyse the even sectors for the two families of solutions. Since transitions of states $ \Omega $ and $ \bar{\Omega} $ are trivial, we focus on the submatrices $A$ and $B$.
\begin{itemize}
	\item Solution $\psi'_l= \alpha_x \psi_{x_l} + \beta\psi_{i_l}^\dag\psi_{i_l}\psi_{x_l}  + \beta\psi_{j_l}^\dag\psi_{j_l}\psi_{x_l}  + \gamma_{ijx}\psi_{i_l}^\dag\psi_{i_l}\psi_{j_l}^\dag\psi_{j_l}\psi_{x_l} $  \\
	For each of the three cases we have a matrix $ \tilde{U}_1 $ consisting in a diagonal matrix having $\{1, 1 + \beta\}$ as eigenvalues, where we set $\alpha_x= 1$ without loss of generality. We can make $1+\beta$ span the whole circumference for different values of $\theta$ and we have perfect discrimination for $ \theta=0$ and $1+\beta=-1$. Indeed, the polygon degenerate in a segment passing for the centre of the circumference.
	\item Solution $\psi'_l= \alpha_x \psi_{x_l} + \alpha_y \psi_{y_l} + \beta_{xy}\psi_{x_l}^\dag\psi_{x_l}\psi_{y_l}  + \beta_{yx}\psi_{y_l}^\dag\psi_{y_l}\psi_{x_l}$ \\
	In this case we deal with two coefficients related to linear operators, $\alpha_x $ and $\alpha_y$ with $x\neq y$ and $x,y \in \{a,b,e\}$. We can easily notice that for every possible choice of indexes $x, y$, the only different elements between $U_1$ and $U_0$ are diagonal for $U_1$ and are equal to $(\alpha_x^*)^2 + (\alpha_y^*)^2 + \beta_{yx}^*\alpha_x^* + \beta_{xy}^*\alpha_y^*$. Consequently, eigenvalues of $\tilde{U}_1$ are $\{1, ((\alpha_x^*)^2+ (\alpha_y^*)^2)((\alpha_x^*)^2+ (\alpha_y^*)^2 + \beta_{yx}^*\alpha_x^* + \beta_{xy}^*\alpha_y^*)\}$. The condition for discriminability is 
	\[ \Phi[((\alpha_x^*)^2+ (\alpha_y^*)^2)((\alpha_x^*)^2+ (\alpha_y^*)^2 + \beta_{yx}^*\alpha_x^* + \beta_{xy}^*\alpha_y^*)]= \pi \] where $\Phi[z]$ is the phase of complex number $z$.
\end{itemize}
\end{appendix}
\end{document}